\documentclass[superscriptaddress,aps,prd,twocolumn,showpacs,nofootinbib,longbibliography, superscriptaddress, showkeys]{revtex4-2}

\DeclareUnicodeCharacter{02BC}{\textendash}
\usepackage[utf8]{inputenc}
\usepackage[T1]{fontenc}
\usepackage{amsmath,amssymb,amsthm}
\usepackage{easybmat}
\usepackage[colorlinks=true,citecolor=blue,urlcolor=blue, linkcolor= magenta]{hyperref}
\usepackage[pdftex]{graphicx}
\usepackage{times, txfonts}
\usepackage{braket}
\usepackage{color}
\usepackage{ragged2e}
\usepackage{booktabs}
\usepackage{epstopdf}
\usepackage{gensymb}
\usepackage{natbib}
\usepackage{subcaption}
\usepackage{float}
\newcommand{\be}{\begin{equation}}
	\newcommand{\ee}{\end{equation}}
\newcommand{\ba}{\begin{eqnarray}}
	\newcommand{\ea}{\end{eqnarray}}

\usepackage{tikz,xcolor,hyperref}

\definecolor{lime}{HTML}{A6CE39}
\DeclareRobustCommand{\orcidicon}{\hspace{-4pt}
	\begin{tikzpicture}
		\draw[lime, fill=lime] (0,0) 
		circle [radius=0.16] 
		node[white] {\hspace{0.1mm}{\fontfamily{qag}\selectfont \tiny ID}};
		\draw[white, fill=white] (-0.07,0.1) 
		circle [radius=0.01];
	\end{tikzpicture}
	\hspace{-3.2mm}
}

\foreach \x in {A, ..., Z}{\expandafter\xdef\csname 
	orcid\x\endcsname{\noexpand\href{https://orcid.org/\csname orcidauthor\x\endcsname}
		{\noexpand\orcidicon}}
}

\begin{document}
 \title{Probing the quantum speed limit and entanglement in flavor oscillations of neutrino-antineutrino system in curved spacetime}
	\author{Abhishek Kumar Jha\orcidA{}}\email{kjabhishek@iisc.ac.in}
\author{Mriganka Dutta\orcidC{}}\email{mrigankad@iisc.ac.in }
\author{Mayank Pathak\orcidD{}}\email{mayankpathak@iisc.ac.in}
\affiliation{Department of Physics, Indian Institute of Science, Bangalore 560012, India}
\author{Subhashish Banerjee\orcidE{}}\email{subhashish@iitj.ac.in}
\affiliation{Indian Institute of Technology Jodhpur, Jodhpur 342011, India}
\author{Banibrata Mukhopadhyay\orcidB{}}\email{bm@iisc.ac.in}
\affiliation{Department of Physics, Indian Institute of Science, Bangalore 560012, India}
	



\begin{abstract}
{We consider a spinning primordial black hole (PBH) described by the Kerr metric in Kerr-Schild polar coordinates. We derive an analytical expression for the four-vector gravitational potential in the underlying Hermitian Dirac Hamiltonian using these coordinates. This gravitational potential introduces an axial vector term in the Dirac equation in curved spacetime. We find that the magnitudes of the temporal and spatial components of the four-vector gravitational potential are significantly affected by the angle of the position vector of the spinor with respect to the spin axis of the PBH, its radial distance from the PBH, and the strength of the specific angular momentum of the PBH. These potentials modify the effective mass matrix of the neutrino-antineutrino system and significantly affect the transition probabilities during the flavor oscillation of the neutrino-antineutrino system. We then use the transition probability to investigate the quantum speed limit time bound ratio for the two-flavor oscillation of the neutrino-antineutrino system in curved spacetime. This helps us estimate how quickly the initial neutrino flavor state evolves over time under the influence of the gravitational field. Finally, we discuss quantum correlations such as entanglement entropy during the two-flavor oscillation of the neutrino-antineutrino system near a spinning PBH.}
\end{abstract}

\maketitle

\section{Introduction}
Neutrino physics plays a crucial role in various astrophysical and cosmological problems, such as neutrino-cooled accretion disks \cite{Chen:2006rra}, r-process nucleosynthesis in supernova explosions \cite{Surman_2008} and the underlying accretion disks around black holes \cite{Mukhopadhyay:1999ut}, leptogenesis-baryogenesis \cite{PhysRevD.45.455} etc. On the other hand, primordial black holes (PBHs) are one of the main candidates for dark matter \cite{Green:2020jor}. The neutrinos being produced by the aforementioned processes interact gravitationally (with the dark matter \cite{Gherghetta:2023myo}). Hence, understanding the dynamics of neutrinos around PBHs can lead to an enhanced understanding of probing dark matter using neutrinos. These studies necessitate the consideration of neutrinos in curved spacetime.

The idea of neutrino oscillations was initially proposed by B. Pontecorvo \cite{Pontecorvo:1957cp, Pontecorvo:1957qd, Bilenky:1978nj, Bilenky:2004xm, https://doi.org/10.1155/2013/873236}, which is experimentally well established in flat spacetime \cite{Bahcall:2004ut,KamLAND:2002uet,KamLAND:2004mhv,Super-Kamiokande:2004orf,MINOS:2006foh,T2K:2013ppw,T2K:2013bzi,Giganti:2017fhf}. Neutrino oscillation is a quantum phenomenon that describes neutrinos transitioning between flavor states due to their nonzero mass \cite{doi:10.1142/5024}. Any initial neutrino flavor state $\nu_\alpha$ ($\alpha=e,\mu,\tau$) can be expressed as a linear superposition of three nondegenerate mass eigenstates $\nu_i$ ($i=1,2,3$). The time evolution of a flavor state is a superposition of the time evolved mass eigenstates.

Flavor oscillations of the neutrino-antineutrino system under gravity have been studied \cite{Mukhopadhyay:2007vca,Sinha:2007uh}, with gravity affecting the neutrino Lagrangian density as $\mathcal{L}=\mathcal{L}_{F}+\mathcal{L}_{I}$. Here, $\mathcal{L}_{F}$ is the free part, and $\mathcal{L}_{I}$ accounts for gravitational interactions, leading to charge-parity-time ($CPT$) violations in stationary (Kerr, etc.) as well as nonstationary (FLRW) spacetimes in local coordinates \cite{Barger:1998xk,BARENBOIM2002227,Mohanty:2002et,Singh:2003sp,Ahluwalia:2004sz,Mukhopadhyay:2005gb,Debnath:2005wk}. In the Weyl representation, $\mathcal{L}_{I}$ introduces gravitational interactions for neutrinos and antineutrinos, altering their dispersion relations. These dispersion relations describe the gravitational Zeeman effect (GZE) \cite{Mukhopadhyay:2007vca,Sinha:2007uh,Barger:1998xk,BARENBOIM2002227,Mohanty:2002et,Singh:2003sp,Ahluwalia:2004sz,Mukhopadhyay:2005gb,Debnath:2005wk,Mukhopadhyay:2021ewd}. This causes flavor oscillations of the neutrino-antineutrino system in curved spacetime. Earlier studies investigated the detailed dynamics of the neutrino-antineutrino system and its associated two-flavor oscillations, specifically using the Kerr-Schild form in Cartesian-type coordinates \cite{Mukhopadhyay:2007vca,Sinha:2007uh,Jha:2024gdq} and the Boyer-Lindquist coordinates \cite{Jha:2024asl}. These studies showed that the influence of $\mathcal{L}_I$ becomes significant only in regions with high spacetime curvature, where the Compton wavelength of the underlying spinors is of the same order as the size of the underlying gravitational source. Due to this, we opt to examine the dynamics of our neutrino-antineutrino system around a spinning PBH.
  
  In this work, we consider the spinning PBH described by the Kerr metric in the fiducial Kerr-Schild polar (KSP) coordinates \cite{Takahashi:2007yi,Misner:1973prb}. 
  We derive an analytical expression for the spin-gravity coupling in the underlying Hermitian Dirac Hamiltonian using these KSP coordinates. This leads to an axial vectorlike gravitational interaction term in the Dirac equation in curved spacetime, contributing to the effective mass matrix of the neutrino-antineutrino systems. This mass matrix is Hermitian and adheres to unitary transformations. We study the unitary evolution of the neutrino-antineutrino system using the modified mass-matrix. From the perspective of an observer at infinity, the magnitudes of the temporal and spatial components of the four-vector gravitational potential affect the transition probabilities of the neutrino-antineutrino system and hence the two-flavor oscillation in the ultrarelativistic limit. The fundamental question arises: how fast does the transition of the two-flavor neutrino-antineutrino state occur with the change in the strength of the four-vector gravitational potential. This idea can be studied by using the concept of the quantum speed limit time (QSLT) \cite{Thakuria:2022taf}. 
  
  The initial discovery of QSLT stems from the uncertainty relationship between conjugate variables in quantum mechanics \cite{Mandelstam1991,MARGOLUS1998188}, which describes the minimum time required for the quantum system to evolve from initial state to final state. The notion of QSLT was investigated for neutral mesons \cite{Banerjee:2022ckm} and electrons in presence of magnetic field \cite{Aggarwal:2021xha}. Recently, it was explored for fermionic particles near Schwarzschild black hole \cite{Maleki:2019cqn}, and extended to neutrino system in flat spacetime \cite{Khan:2021kai,Bouri:2024kcl}. In addition, the influence of gravity on the QSLT in neutrino oscillations was explored, specifically using the Kerr metric in Cartesian-type coordinates \cite{Jha:2024gdq} and in Boyer-Lindquist coordinates \cite{Jha:2024asl}.  In the context of two-flavor oscillations of the neutrino-antineutrino system in curved spacetime, the QSLT can be understood as the maximal rate at which neutrino flavor changes, and it provides a lower bound on the time required to change flavor. This lower bound is referred as the QSLT, which is defined as the ratio of distinguishability between flavor state to the energy fluctuation. The distinguishability between the initial and final states is quantified by the Bures angle, which is related to the transition probability of the evolved flavor neutrino system. In this work, we use the QSLT bound ratio technique as the primary analytical tool to examine the two-flavor oscillation of the neutrino-antineutrino system in curved spacetime, employing the Kerr metric parameters in KSP coordinates.

Moreover, entanglement is a prominent example of quantum correlations, which are fundamental aspects of quantum information theory and exhibit intriguing features that have no classical counterparts \cite{Nielsen:2012yss}. The study of quantum correlations is crucial for understanding and applying quantum mechanics in various natural systems. Although much research has been focused on optical and electronic systems, recent advances in neutrino oscillation experiments in flat space-time offer new avenues for exploring various quantum correlations in oscillating neutrinos \cite{Blasone_2009,Blasone:2007wp,Banerjee:2015mha,ALOK201665,doi:10.1142/S0217732321500565,Jha:2021itm,Jha:2020qea,Jha:2022yik,Bittencourt:2022tcl,Banerjee:2024lih}.
In these studies, occupation numbers are defined within a multiqubit space, allowing the neutrino flavor state to be interpreted as a multipartite mode-entangled state. Neutrinos interact weakly; thus, a neutrino beam can be coherent over long distances, which may have implications in quantum information theory. Motivated by these studies, Refs.\cite{Dixit:2019lsg,Mukhopadhyay:2018oli} rekindled interest in quantifying various quantum correlations, including von Neumann entropy and nonclassicality features such as the Mermin and Svetlichny inequalities, etc., in the framework of two-flavor oscillations of the neutrino-antineutrino system in curved spacetime. Their results emphasize the implication of entanglement in gravity-induced flavor oscillations of the neutrino-antineutrino system. In quantum information theory \cite{Nielsen:2012yss,Guhne:2008qic}, the most basic type of multipartite quantum system is a bipartite quantum system, with Bell's state being a prominent example. In the present work, we consider the two-flavor oscillation of the neutrino-antineutrino system in curved spacetime as a bipartite quantum system. We delve into the behavior of quantum correlations such as the entanglement entropy for the two-flavor oscillation of the neutrino-antineutrino system, particularly near a spinning PBH.

The organization of the paper is as follows. In Sec.\,\ref{sec2}, we provide a brief review of the QSLT for unitary quantum evolution. Sections\,\ref{sec3} and \ref{sec4} discuss the Lagrangian formulation in curved spacetime and derivations of four-vector gravitational potential using  KSP coordinates, respectively. Section\,\ref{sec5} describes the formalism of the effective mass matrix in the neutrino-antineutrino system. Section\,\ref{sec6} investigates oscillations in the neutrino-antineutrino system within curved spacetime. Subsequently, Sec.\,\ref{sec7} delves into two-flavor oscillations of the neutrino-antineutrino system in curved spacetime. Section\,\ref{sec8} is dedicated to exploring the QSLT for the two-flavor oscillation of the neutrino-antineutrino system in the spacetime of the spinning PBH. In Sec.\,\ref{sec9}, we discuss the implications of entanglement in the two-flavor oscillation of the neutrino-antineutrino system in the presence of a gravitational field. Finally, Sec.\,\ref{sec10} presents our conclusions.

\section{Quantum speed limit time (QSLT) for unitary quantum evolution}
\label{sec2}
In quantum mechanics, the wave function $\ket{\nu}$ describes the state of a quantum system, and its dynamics govern the temporal evolution of this state according to the Schrödinger equation as
  \begin{eqnarray}
      i\hbar\frac{d\ket{\nu}}{dt}=\mathcal{H}\ket{\nu}\Longrightarrow\ket{\nu(t)}=e^{-i \mathcal{H} t/\hbar}\ket{\nu(0)}\equiv \mathcal{U}_t\ket{\nu(0)},
      \label{a.1}
\end{eqnarray}
   with $\mathcal{U}_t$ being the unitary time evolution operator, and  $\mathcal{H}$ is Hermitian ($\mathcal{H}^\dagger=\mathcal{H}$) and time independent driving Hamiltonian. QSLT ($\mathcal{T}_{\text{QSLT}}$) for such a system is defined as \cite{Thakuria:2022taf,Mandelstam1991}
\begin{align}
\mathcal{T}_{\text{QSLT}}= \frac{\hbar \mathcal{S}_0}{ \Delta {{\mathcal{H}}}}\hspace{0.2cm} \text{and}\hspace{0.2cm} \frac{\mathcal{T}_{\text{QSLT}}}{\mathcal{T}}\leq 1
\label{a.2}
 \end{align}
is the QSLT bound ratio, where $ \mathcal{\mathcal{T}}$ is the propagation time of the initial state. Here,
\begin{equation}
    {\mathcal{S}_0} = \cos^{-1}(|\langle \nu (0)|\nu(\mathcal{T})\rangle|)=\arccos(\sqrt{P_s})
    \label{a.3}
\end{equation}
is the Bures angles (or geodesic distance) between initial and final states, and $P_s=|\langle \nu (0)|\nu(\mathcal{T})\rangle|^2$ is the survival probability of the initial state $\ket{\nu}$ at the propagation time $\mathcal{\mathcal{T}}$. A general representation of the energy fluctuation $\Delta \mathcal{H}$ for a quantum system undergoing unitary evolution is given by \cite{Thakuria:2022taf}
\begin{equation}
   \Delta{\mathcal{H}}\equiv \sqrt{ [\langle \dot{\nu} (t)|\dot{\nu}(t)\rangle-(i\langle \nu (t)|\dot{\nu}(t)\rangle)^2 ]},
    \label{a.4}
\end{equation}
where ${\ket{\dot{\nu}(t)}}\equiv d\ket{\nu(t)}/dt$. The physical interpretation of $\mathcal{T}_{\text{QSLT}}$ is as follows:
\begin{itemize}
    \item If the QSLT bound ratio ${\mathcal{T}_{\text{QSLT}}}/{\mathcal{T}}=1$, the quantum state's dynamic evolution cannot be further speed up. In other words, the evolution speed is already at its maximum.
    \item If the QSLT bound ratio ${\mathcal{T}_{\text{QSLT}}}/{\mathcal{T}}<1$, the dynamic evolution of the quantum state could be speed up. Furthermore, the smaller ${\mathcal{T}_{\text{QSLT}}}/{\mathcal{T}}$ is, the larger is the potential to quantum speed up.
\end{itemize}

\section{Lagrangian Formulation}
\label{sec3}
In the curved spacetime, the Dirac Lagrangian for spin-1/2 particles gets modified \cite{Mukhopadhyay:2007vca,Sinha:2007uh,Schwinger:2019lox}. Let us briefly discuss the coupling of gravity, to the quantum particle, by the following Lagrangian density
\begin{equation}
    {\mathcal{L}}=\sqrt{-g}~(\Bar{\Psi}i\gamma^{\mu}\overset\longleftrightarrow{D_{\mu}}~\Psi-\Bar{\Psi}m\Psi),
    \label{1.1}
\end{equation}
where $D_{\mu}$ is the covariant derivative defined as
\begin{equation}
    D_{\mu}=\partial_{\mu}-\frac{i}{4}\omega_{\nu \lambda \mu }\sigma^{\nu \lambda }.
    \label{1.2}
\end{equation}
The above Lagrangian is valid for curved as well as locally flat spacetimes because the term $\gamma^{\mu}\overset\longleftrightarrow{D_{\mu}}$ remains invariant. Therefore, we can write it as $\gamma^{c}\overset\longleftrightarrow{D_{c}}$\footnote{Throughout the paper, Greek indices mean global coordinate system and Roman indices mean local coordinate system.}. The spin connections in the covariant derivative $D_c$ are defined as
\begin{equation}
    \omega_{bac} = e_{b\lambda} (\partial_c e^{\lambda}_{a} + \Gamma^{\lambda}_{\alpha\mu}e^{\alpha}_{a}e^{\mu}_{c}),
    \label{1.3}
\end{equation}
and
\begin{equation}
    \sigma^{ba}=\frac{i}{2}[\gamma^{b},\gamma^{a}].
    \label{1.4}
\end{equation}
Here, the $e$'s are the tetrads which connect between local and global spacetimes. The equation of motion for $\Psi$ will then be 
\begin{equation}
    (i\gamma^{c}\partial_{c} - m)\Psi - \frac{i}{8} \gamma^{c} \omega_{bac} [\gamma^{b},\gamma^{a}] \Psi=0.
    \label{1.5}
\end{equation}
The second term in the equation can be expanded as 
\begin{equation}
\frac{i}{4} \omega_{bac}[\eta^{cb}\gamma^a-\eta^{ac}\gamma^b-i\gamma_d \epsilon^{dcab}\gamma^5],
\label{1.6a}
\end{equation}
which we define as $iA_a\gamma^a-B^d\gamma_d\gamma^5$. Therefore,
\begin{equation}
B^d=\epsilon^{abcd}\omega_{bac},
\label{1.7}
\end{equation}
is the four-vector gravitational potential. The term consisting of $A_a$ cancels out, in the Lagrangian itself, when we take the contribution of the covariant derivative on the $\Bar{\Psi}$, i.e. $\Psi~i\gamma^{\mu}\overset\longleftrightarrow{D_{\mu}}~\Bar{\Psi}$. Thus, the Lagrangian in curved spacetime can be simplified as
\begin{equation}
   (-g)^{-1/2} \mathcal{L}=\bar{\Psi}(i~\gamma^c\overset\longleftrightarrow{\partial_{c}}-m+ B^d\gamma_d\gamma^5)\Psi.
    \label{1.8}
\end{equation}
The Lagrangian given by Eq.\,(\ref{1.8}) consists of two terms: the free term and the axial vector interaction term. The interaction term couples to the field $B^d$, which remains constant in a local inertial frame due to its origin in the background gravitational field.

\section{Gravitational Potential around a spinning Black Hole}
\label{sec4}
In the Kerr-Schild polar (KSP) coordinates the Kerr metric is \cite{Takahashi:2007yi}
\begin{align}
    ds^2&=-(1-2Mr/\rho^2)\mathrm{d}t^2+4Mr/\rho^2\mathrm{d}r\mathrm{d}t-(4Mar\sin^2\theta/\rho^2)\mathrm{d}t\nonumber\\&\mathrm{d}\phi+(1+2Mr/\rho^2)\mathrm{d}r^2+\rho^2 \mathrm{d}\theta^2-2a\sin^2\theta(1+2Mr/\rho^2)\mathrm{d}r\nonumber\\&\mathrm{d}\phi+A\sin^2\theta/\rho^2\mathrm{d}\phi^2,
    \label{1}
\end{align}
where $\rho^2=r^2+a^2 \cos^2\theta$, $A=(r^2+a^2)^2-a^2\Delta\sin^2\theta$, $\Delta=r^2-2Mr+a^2$, $M$ is the mass of black hole (BH), $a$ is the dimension less specific angular momentum per unit mass ($-1\leq a\leq 1$), and $\theta\in [0,\pi]$ is the angle of the position vector of the spinor\footnote{In this work, we choose the neutrino spinor, which will be discussed in the next section.} with
respect to the spin axis of the BH. Here, we choose $GM=c=\hbar=1$, where $G$ is the gravitational constant, $\hbar$ is the reduced Planck constant, and $c$ is the speed of light. We measure distance in units of the gravitational radius of the BH as $r_g=GM/c^2=1$. The outer event horizon of the rotating BH is at (in units of $r_g$), 
\begin{equation}
    r_{+}=1 + \sqrt{1 - a^2}.
    \label{2}
\end{equation}
Without loss of generality, we choose the nonzero tetrads 
as \cite{Takahashi:2007yi} 
\begin{align}
  e^t_{~0}&=(1+2Mr/\rho^2)^{1/2}, ~~~e^r_{~0}=-2Mr/(\rho^2(\rho^2+2Mr))^{1/2},\nonumber\\
 & e^r_{~1}=[A/(\rho^2(\rho^2+2Mr))]^{1/2},~~~e^{\theta}_{~2}=1/\rho,\nonumber\\
  &e^{\phi}_{~1}=a((\rho^2+2Mr)/\rho^2A)^{1/2},~~~e^{\phi}_{~4}=(\rho^2/A)^{1/2}/\sin{\theta}.
    \label{3}
\end{align}

\begin{figure}[!htbp]
   \centering
    \begin{subfigure}{0.99\columnwidth}
       \centering
       \caption{} \label{Fig2a}
       \includegraphics[width=\textwidth]{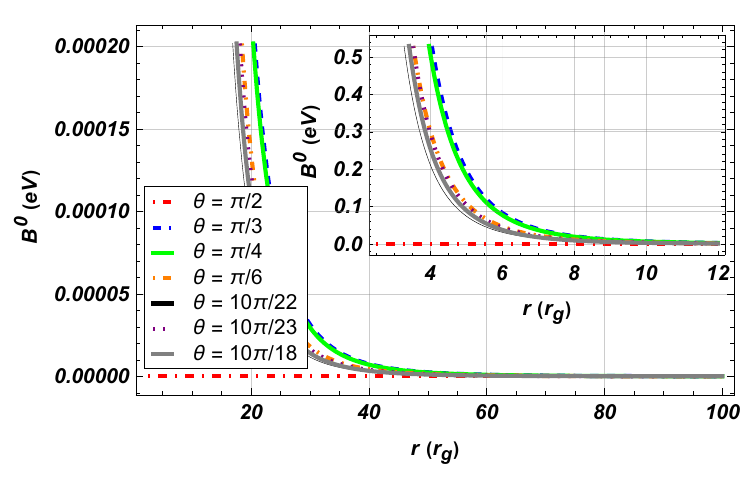}
    \end{subfigure}
   \hfill
   \begin{subfigure}{0.99\columnwidth}
        \centering
         \caption{}\label{Fig2b}
        \includegraphics[width=\textwidth]{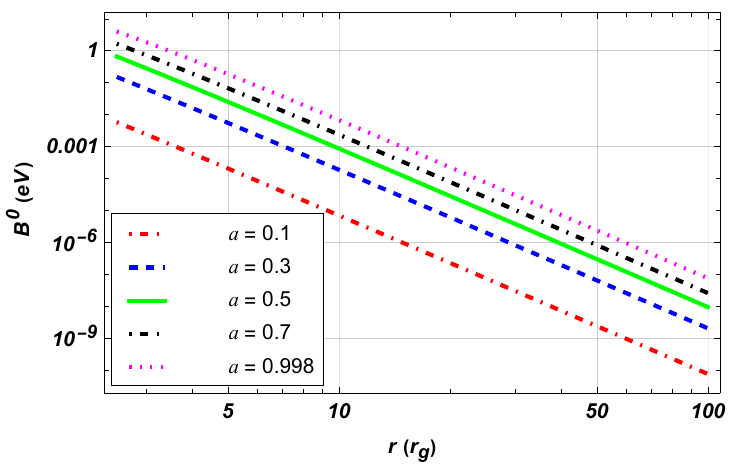}  
    \end{subfigure}
   \caption{\justifying{Magnitude of the temporal gravitational potential $B^0$ as a function of the radial distance $r$: (a) in the top panel, for different values of angle $\theta$, with $a=0.998$; and (b) in the bottom panel, for different values of $a$,
    with $\theta=\pi/4$, respectively.}}
    \label{Fig2}
\end{figure}

\begin{figure}[!htbp]
   \centering
    \begin{subfigure}{0.99\columnwidth}
       \centering
       \caption{}
        \label{Fig3a}
       \includegraphics[width=\textwidth]{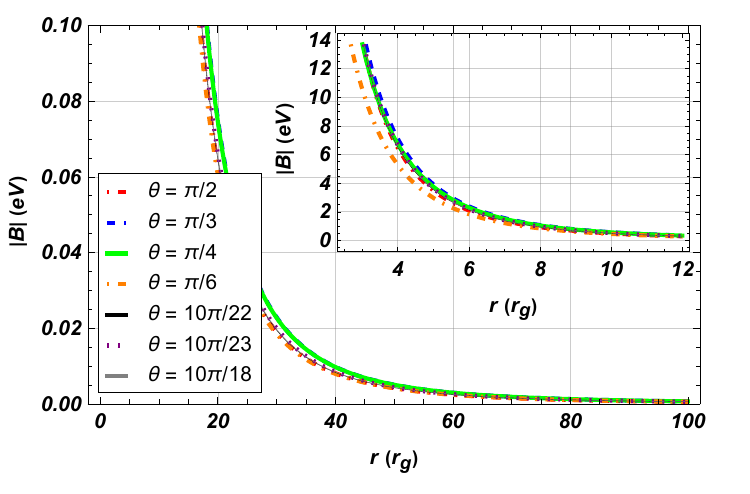}
    \end{subfigure}
   \hfill
   \begin{subfigure}{0.99\columnwidth}
        \centering
         \caption{}
        \label{Fig3b}
        \includegraphics[width=\textwidth]{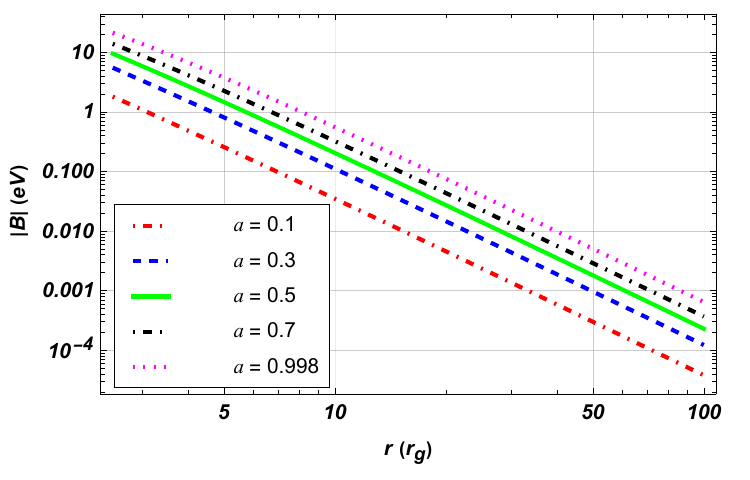}
    \end{subfigure}
   \caption{\justifying{Same as Fig.\,\ref{Fig2}, except $|\textbf{B}|$ is shown.}}
    \label{Fig3}
\end{figure}

\begin{figure}[!htbp]
   \centering
    \begin{subfigure}{0.99\columnwidth}
       \centering
       \caption{}\label{Fig4a}
       \includegraphics[width=\textwidth]{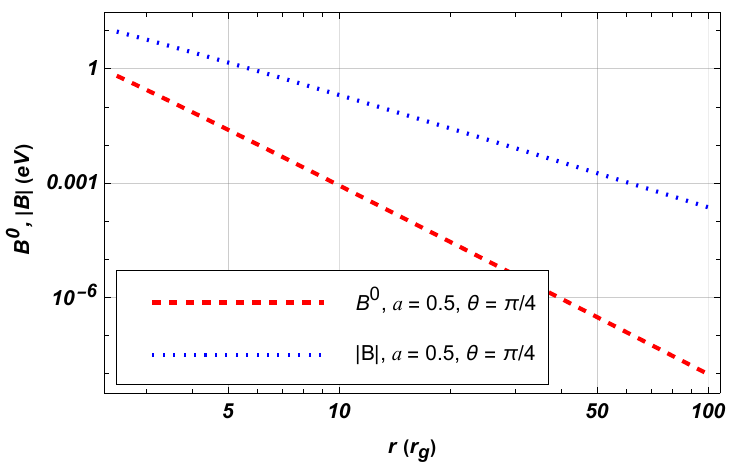}
    \end{subfigure}
   \hfill
   \begin{subfigure}{0.99\columnwidth}
        \centering
        \caption{}
        \label{Fig4b}
        \includegraphics[width=\textwidth]{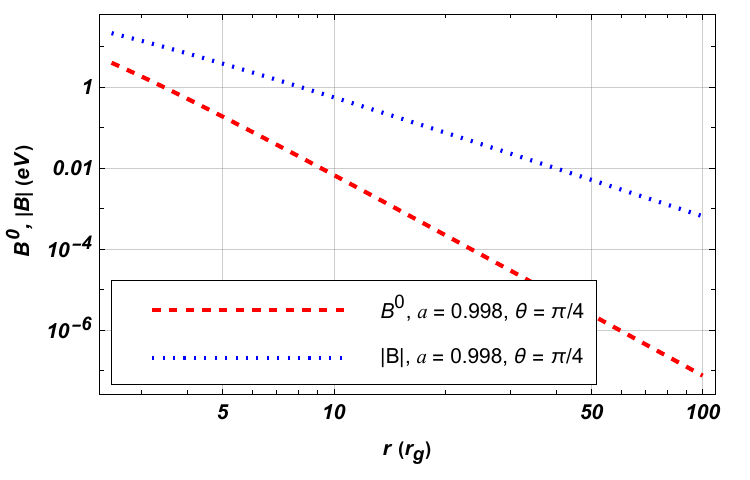}
    \end{subfigure}
   \caption{\justifying{Comparison between $B^0(\text{eV})$ and $|\textbf{B}|(\text{eV})$ as functions of the radial distance $r$ is shown at $\theta=\pi/4$: (a) in the top panel, for a low value of $a=0.5$; and (b) in the bottom panel, for a high value of $a=0.998$.}}
    \label{Fig4}
\end{figure}

Now, we have to find the expression for $B^d$ given by Eq.\,(\ref{1.7}). To do so, one needs to estimate the mass of BH. The effect of gravity is dominated by the quantum nature of neutrinos when the gravitational radius ($r_g$) of the BH is comparable with the Compton wavelength of neutrinos ($\lambda_c$), i.e. $r_g \sim \lambda_c$. We take the mass of the neutrino in the order of $5\times 10^{-2}$ eV, which essentially sets the mass of PBH $\leq$ $10^{21}$ \text{kg}.
Throughout this work, we take the mass of the PBH as $M_{\text{PBH}}=10^{18}$ \text{kg} (unless mentioned otherwise), which turns out to be $5.61\times10^{53} \text{eV}$. The four-vector gravitational potential $B^d(\rm{eV})$ can be calculated in the unit of gravitational radius, $r_{\textit{g}}(\rm{{eV}^{-1}})$ as
\begin{equation}
B^d=\left[\epsilon^{abcd}e_{b\lambda}\left(\partial_a e_c^{\lambda} + \Gamma_{\alpha\mu}^{\lambda} e_c^{\alpha}e_a^{\mu}\right)\right](\hbar c/r_g).
\label{4}
\end{equation}
It is important to note that when the spin parameter $a=0$, all components of $B^d$ vanish. 
 
Figure\,\ref{Fig2} depicts the temporal component of $B^d$,
$B^0$(\text{eV}), as a function of the radial distance $r$. In the top panel of Fig.\,\ref{Fig2}, $B^0$ vs $r$ is shown for different values of $\theta$, with a high value of the specific angular momentum, $a=0.998$. We find that at $\theta=\pi/2$ (red dot-dashed line), $B^0$ tends to zero. However, for other values of $\theta$, $B^0$ remains nonzero. The strength of $B^0$ is highest near the PBH at $\theta=\pi/3$ (blue dashed line), and lowest at $\theta=\pi/2$ (red dot-dashed line), for the different values of $\theta$ depicted. The zoomed-in figure near the PBH further validates this result. Moreover, the strength of $B^0$ decreases to zero as the distance from the PBH increases. In the bottom panel of Fig.\,\ref{Fig2}, $B^0$ vs $r$ is shown for different values of $a$, with a fixed angle $\theta=\pi/4$. We observe that $B^0$ decreases monotonically with distance from the PBH for all values of $a$. However, for larger values of specific angular momentum, such as $a=0.998$ (magenta dotted line), the magnitude of $B^0$ remains higher than the other black hole spins considered, throughout the entire range of $r$.

Furthermore, by Eq.\,(\ref{4}), we determine the values of three spatial components of $B^d$ : $B^1$, $B^2$, and $B^3$ in units of \text{eV}. Particularly, at $\theta=\pi/2$, we find that the magnitudes of $B^1$ and $B^3$ are zero, while the magnitude of $B^2$ is nonzero. At other values of $\theta\in [0,\pi]$, the magnitudes of all three individual spatial components of $B^d$ are nonzero. The magnitude of the spatial part of the four-vector gravitational potential can be written as 
\begin{equation}
|\textbf{B}|=\sqrt{({B^1})^2+({B^2})^2+(B^3)^2}\rm\,{eV}.
\end{equation}
In the top and bottom panels of Fig.\,\ref{Fig3}, we illustrate $|\textbf{B}|\rm\,{(eV)}$ as a function of $r$ for different values of $\theta$ with fixed $a=0.998$ and different values of $a$ with fixed $\theta=\pi/4$, respectively. Its behavior looks similar to $B^0$, except for $\theta=\pi/2$, due to the non-zero value of $B^2$.

Moreover, the curvature of spacetime for a rotating BH depends fundamentally on its mass and spin. A spinning BH will have higher spacetime curvature near it, compared to a nonspinning BH of the same mass. The higher curvature provides a higher gravitational potential, which can also be confirmed from $B^0$ (Fig.\,{\ref{Fig2}}) and $|\textbf{B}|$ (Fig.\,{\ref{Fig3}}).

Furthermore, in the top and bottom panels of Fig.\,{\ref{Fig4}}, we compare $B^0$ (red dashed line) and $|\textbf{B}|$ (blue dotted line) as functions of $r$, with $\theta=\pi/4$, for a low  $a=0.5$ and a high $a=0.998$, respectively.  We observe that the strength of $|\textbf{B}|$ (blue dotted line) remains higher than that of $B^0$ (red dashed line) for both low and high values of $a$.

Thus, these results demonstrate that the strength of the four-vector gravitational potential ($B^d$) is highly dependent on $\theta$ of the position vector of the spinor with respect to the spin axis of the PBH, apart from the radial distance from the PBH ($r$), and $a$.

\section{Modified neutrino mass matrix}
\label{sec5}
We consider neutrino to be a left-handed particle. For Majorana neutrinos, neutrino spinor in the Weyl representation can be expressed as \cite{Sinha:2007uh}
\begin{equation}
    \ket{\Psi}=\begin{pmatrix}
        \ket{\psi^c}\\
        \ket{\psi}
    \end{pmatrix},
    \label{4.1}
\end{equation}
where $\ket{\psi^c}$ and $\ket{\psi}$ denote antineutrino and neutrino spinors, having lepton number eigenvalues -1 and 1, respectively. The Majorana mass ($m$) in terms of two-component spinors can be expressed as
\begin{equation}
    \bar{\Psi}\mathcal{M}\Psi=\begin{pmatrix}
        {\psi^c}^\dagger & \psi^\dagger
    \end{pmatrix}\begin{pmatrix}
        0 & -m\\
        -m & 0
    \end{pmatrix}\begin{pmatrix}
        \psi^c\\
        \psi
    \end{pmatrix}.
    \label{4.2}
\end{equation}
Following Eqs.\,(\ref{1.8}) and (\ref{4}), the Lagrangian density in the presence of the gravitational field can be expanded as
\begin{align}
    (-g)^{-1/2}\mathcal{L}&=\begin{pmatrix}
        {\psi^c}^{\dagger} & {\psi}^{\dagger}
    \end{pmatrix}{i}\gamma^0\gamma^c\mathcal{\overleftrightarrow{\partial}}_c\begin{pmatrix}
        {\psi^c}\\ {\psi}
    \end{pmatrix}+\begin{pmatrix}
        {\psi^c}^{\dagger} & {\psi}^{\dagger}
\end{pmatrix}\gamma^0 B^0 \gamma_0 \gamma^5\begin{pmatrix}
        {\psi^c}\\ {\psi}
    \end{pmatrix}\nonumber\\
   & +\begin{pmatrix}
        {\psi^c}^{\dagger} & {\psi}^{\dagger}
\end{pmatrix}\gamma^0 B^1 \gamma_1 \gamma^5\begin{pmatrix}
        {\psi^c}\\ {\psi}
    \end{pmatrix}+\begin{pmatrix}
        {\psi^c}^{\dagger} & {\psi}^{\dagger}
\end{pmatrix}\gamma^0 B^2 \gamma_2 \gamma^5\begin{pmatrix}
        {\psi^c}\\ {\psi}
    \end{pmatrix}\nonumber\\
   & +\begin{pmatrix}
        {\psi^c}^{\dagger} & {\psi}^{\dagger}
\end{pmatrix}\gamma^0 B^3 \gamma_3 \gamma^5\begin{pmatrix}
        {\psi^c}\\ {\psi}
    \end{pmatrix}-\begin{pmatrix}
        {\psi^c}^{\dagger} & {\psi}^\dagger  \end{pmatrix}\gamma^0m\begin{pmatrix}
        {\psi^c}\\ {\psi}
    \end{pmatrix}.
    \label{4.3}
\end{align}
In the Weyl (chiral) basis (alternate form), the gamma matrices can be represented as
\begin{equation}
    \gamma^0=\begin{pmatrix}
        0&-I_2\\
        -I_2 & 0
    \end{pmatrix},\hspace{0.3cm} \gamma^i=\begin{pmatrix}
        0&\sigma^i\\
        -\sigma^i & 0
    \end{pmatrix}, \hspace{0.3cm} \gamma^5=\begin{pmatrix}
        I_2 &0\\
        0& -I_2
    \end{pmatrix},
    \label{4.4}
\end{equation}
where $I_2$ is a $2\times 2$ identity matrix and $\sigma^i$ are Pauli matrices ($i=1,2,3$). Subsequently, using Eq.\,(\ref{4.4}) in Eq.\,(\ref{4.3}), we obtain
\begin{align}
    (-g)^{-1/2}\mathcal{L}=\begin{pmatrix}
        {\psi^c}^{\dagger} & {\psi}^{\dagger}
    \end{pmatrix}{i}\gamma^0\gamma^c\mathcal{\overleftrightarrow{\partial}}_c\begin{pmatrix}
        {\psi^c}\\ {\psi}
    \end{pmatrix}&\nonumber\\-\begin{pmatrix}
        {\psi^c}^{\dagger} & {\psi}^{\dagger}
\end{pmatrix}\begin{pmatrix}
    -B^0- B^i\sigma_i & 0\\
    0 & B^0- B^i\sigma_i
\end{pmatrix}\begin{pmatrix}
        {\psi^c}\\ {\psi}
    \end{pmatrix}&\nonumber\\
    -\begin{pmatrix}
        {\psi^c}^{\dagger} & {\psi}^\dagger  \end{pmatrix}\begin{pmatrix}
            0 & -m\\
            -m & 0
        \end{pmatrix}\begin{pmatrix}
        {\psi^c}\\ {\psi}
    \end{pmatrix}.
    \label{4.5}
\end{align}
In the above equation, the two temporal terms: $B^0{\psi^c}^\dagger\psi^c$ and $B^0{\psi}^\dagger\psi$; and six spatial terms: $B^1\sigma_1{\psi^c}^\dagger\psi^c$, $B^1\sigma_1{\psi}^\dagger\psi$, $B^2\sigma_2{\psi^c}^\dagger\psi^c$, $B^2\sigma_2{\psi}^\dagger\psi$, $B^3\sigma_3{\psi^c}^\dagger\psi^c$, and $B^3\sigma_3{\psi}^\dagger\psi$, do not violate the lepton number and effectively contribute to the $\it{mass}$ of a Majorana neutrino. Consequently, the Lagrangian, in the presence of a background gravitational field, can be written with the effective $\it{mass}$ terms as follows:
\begin{equation}
   i \gamma^0\gamma^c\mathcal{\overleftrightarrow{\partial}}_c\begin{pmatrix}
        {\psi^c}\\ {\psi} \end{pmatrix}- \begin{pmatrix}
   -B^0- B^i\sigma_i & -m\\
    -m & B^0- B^i\sigma_i
\end{pmatrix}\begin{pmatrix}
        {\psi^c}\\ {\psi}
    \end{pmatrix}=0.
    \label{4.6}
\end{equation}
Therefore, the effective mass matrix of the neutrino-antineutrino system in the presence of gravitational field can be obtained as
\begin{equation}
    \mathcal{M}=\begin{pmatrix}
    -B^0-B^1\sigma_1-B^2\sigma_2-B^3\sigma_3 & -m\\
    -m &  B^0-B^1\sigma_1-B^2\sigma_2-B^3\sigma_3\\
  \end{pmatrix}.
  \label{4.6}
\end{equation}
Hence, if one assumes that neutrinos solely possess Majorana-type masses, they acquire effectively lepton number nonviolating masses in addition. Moreover, in this scenario, $\psi$ is not a mass eigenstate.

\section{Oscillation in Neutrino-antineutrino system}
\label{sec6}
\begin{figure}[!htbp]
   \centering
    \begin{subfigure}{0.99\columnwidth}
       \centering
       \caption{}
        \label{Fig5a}
       \includegraphics[width=\textwidth]{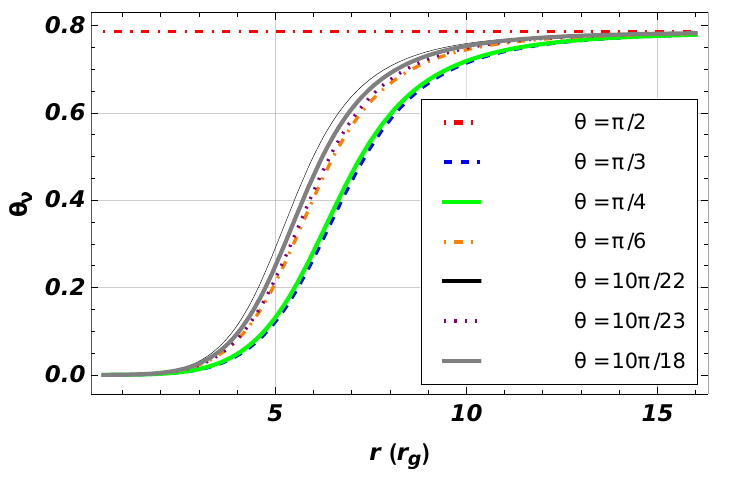}
    \end{subfigure}
   \hfill
   \begin{subfigure}{0.99\columnwidth}
        \centering
        \caption{}
        \label{Fig5b}
        \includegraphics[width=\textwidth]{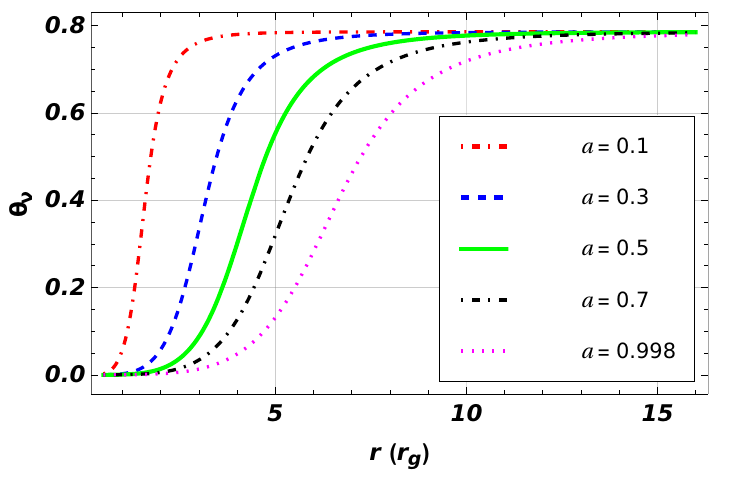}
    \end{subfigure}
   \caption{\justifying{Mixing angle $\theta_\nu$ as a function of the radial distance $r$: (a) in the top panel, for different values of angle $\theta$, with $a=0.998$; and (b) in the bottom panel, for different values of $a$, with $\theta=\pi/4$, respectively.}}
    \label{Fig5}
\end{figure}
The mass matrix $\mathcal{M}$, as given by Eq.\,(\ref{4.6}), is Hermitian and can be diagonalized by a unitary transformation with the operator $\mathbf{U}(\theta_\nu)$. Inspired by the nature of the neutral-kaon system, the two-component spinors $\psi^c$ and $\psi$ are not mass eigenstates but the linear superposition of them. Thus,
\begin{equation}
\begin{pmatrix}
    \ket{\nu_1}\\
    \ket{\nu_2}
\end{pmatrix}=\mathbf{U}(\theta_\nu)\begin{pmatrix}
    \ket{\psi^c}\\
    \ket{\psi}
\end{pmatrix};\hspace{0.3cm} \mathbf{U}(\theta_\nu)=\begin{pmatrix}
    \cos\theta_\nu & \sin\theta_\nu\\
    -\sin\theta_\nu & \cos\theta_\nu
\end{pmatrix},
\label{5.1}
    \end{equation}
where $\ket{\nu_1}$ and $\ket{\nu_2}$ are two mass eigenstates and $\textbf{U}^\dagger \textbf{U}=1$. The mixing angle between neutrino and antineutrino is given by
\begin{equation}
    \theta_\nu=\tan^{-1}\left[\frac{m}{B^0+\sqrt{(B^0)^2+m^2}}\right].
    \label{5.2}
\end{equation}

\begin{figure}[!htbp]
   \centering
    \begin{subfigure}{0.99\columnwidth}
       \centering
        \caption{}
        \label{Fig6a}
       \includegraphics[width=\textwidth]{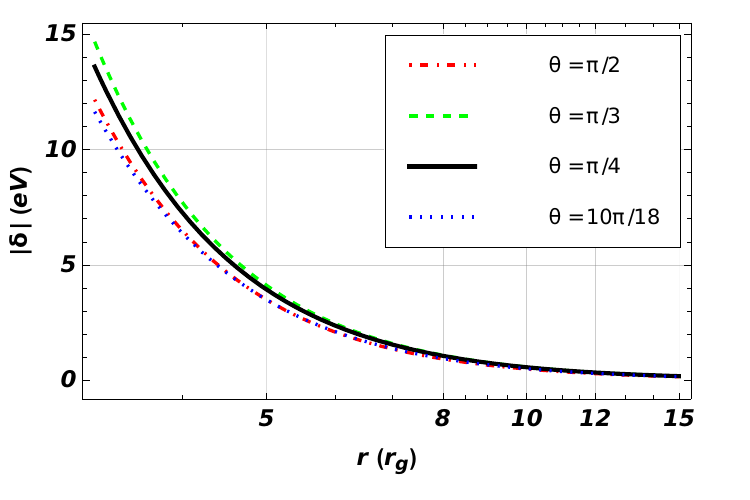}
    \end{subfigure}
   \hfill
   \begin{subfigure}{0.99\columnwidth}
        \centering
         \caption{}
        \label{Fig6b}
        \includegraphics[width=\textwidth]{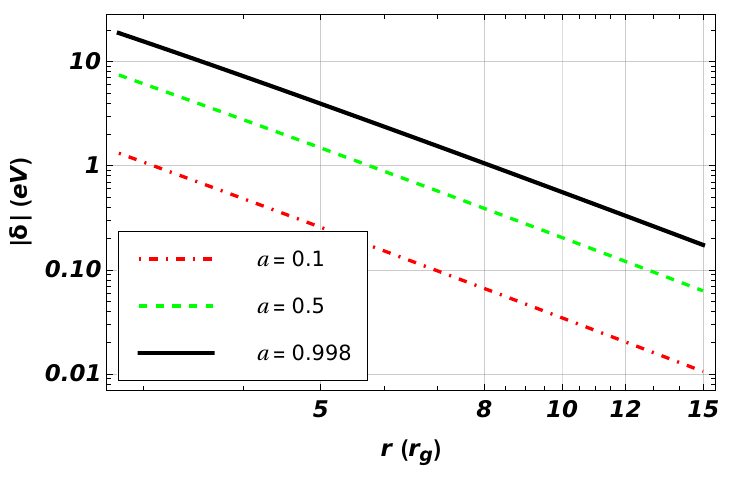}
    \end{subfigure}
   \caption{\justifying{$|\delta|\,(\text{eV})$ (where, $\delta\approx B^0-|\textbf{B}|$) as a function of the radial distance $r$: (a) in the top panel, for different values of angle $\theta$, with $a=0.998$; and (b) in the bottom panel, for different values of $a$, with $\theta=\pi/4$, respectively.}}
    \label{Fig6}
\end{figure}

\begin{figure}[!htbp]
   \centering
    \begin{subfigure}{0.99\columnwidth}
       \centering
       \caption{}
        \label{Fig7a}
       \includegraphics[width=\textwidth]{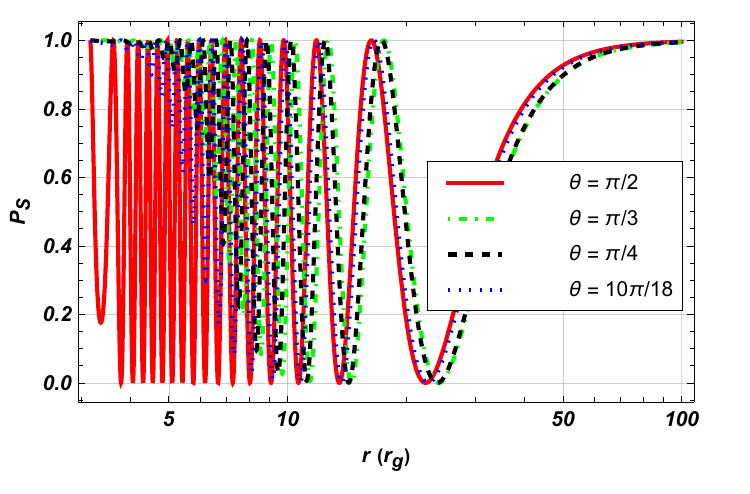}
    \end{subfigure}
   \hfill
   \begin{subfigure}{0.99\columnwidth}
        \centering
        \caption{}
        \label{Fig7b}
        \includegraphics[width=\textwidth]{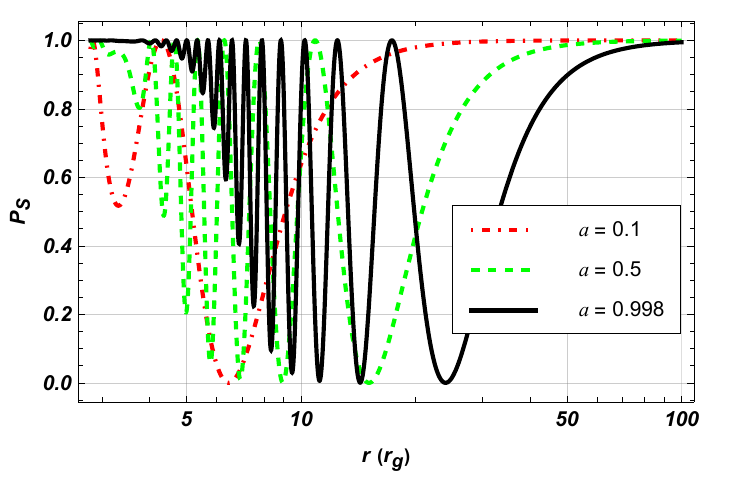}
    \end{subfigure}
   \caption{\justifying{Survival probability $P_S$, of the initial state $\ket{\nu_1}$, as a function of the radial distance $r$: (a) in the top panel, for different values of angle $\theta$, with $a=0.998$; and (b) in the bottom panel, for different values of $a$, with $\theta=\pi/4$, respectively.}}
    \label{Fig7}
\end{figure}

Here, when $m=0$, there is no mixing between $\psi^c$ and $\psi$. Moreover, in the absence of interaction with gravity (i.e., when $B^0=0$), the neutrino-antineutrino mixing angle $\theta_\nu$ is $\pi/4$. A nonzero gravitational field alters the neutrino-antineutrino state from passive (constant $\theta_\nu$) to active (varying $\theta_\nu$). In the top panel of Fig.\,\ref{Fig5}, we illustrate the mixing angle $\theta_\nu$ as a function of the radial distance $r$ for different values of $\theta$, with a fixed $a=0.998$ and Majorana mass\footnote{The Majorana mass given here is calculated in the next section.} $m=0.0497666\rm\,{eV}$. As depicted in the top panel of Fig.\,{\ref{Fig2}}, at a small radial distance $r$ from the PBH, the strength of $B^0$ is very high ($B^0>>m$), which causes $\theta_\nu$ in Fig.\,{\ref{Fig5}} to approach low values. At a large radial distance $r$ from the PBH however, where the strength of $B^0$ is low ($B^0<<m$), $\theta_\nu$ is $\pi/4$ for all values of $\theta$. Since $B^0$ is zero at $\theta=\pi/2$ (see red dot-dashed line in the top panel of Fig.\,\ref{Fig2}), $\theta_\nu$ saturates to $\pi/4$ (see red dot-dashed line in the top panel of Fig.\,\ref{Fig5}), which corresponds to a maximal mixing angle. It can also be seen that for low value of $r$, i.e. near to the PBH, the value of $\theta_\nu$ is small (except for $\theta=\pi/2$). This suggests that oscillations in the neutrino-antineutrino system in this region should be low. Furthermore, in the bottom panel of Fig.\,\ref{Fig5}, we illustrate $\theta_\nu$ variation with $r$ for different values of $a$, with a fixed $\theta=\pi/4$ and $m=0.0497666\rm\,{eV}$. It is observed that initially, the variation of $\theta_\nu$ is sharp for small values of $a$ (red dot-dashed line) and becomes shallower with the increase of $a$. 

Moreover, neutrino and antineutrino states couple together with modified energy
\begin{align}
E_\nu&=\sqrt{({\textbf{p}}-\textbf{B})^2+m^2}+B^0,\hspace{0.5cm}
E_{\nu^c}=\sqrt{(\textbf{p}+\textbf{B})^2+m^2}-B^0,
\label{5.3}
  \end{align}
with momentum $\textbf{p}$. This energy splitting between neutrino and antineutrino states effectively denotes gravitational Zeeman effect (GZE), which leads to possible oscillations in the neutrino-antineutrino system. Due to the difference in energy (or effective mass), the time evolution of neutrino and antineutrino states is distinct, such that
\begin{equation}
\ket{\psi(t)}\rightarrow\ket{\psi(0)}e^{-iE_{\nu} t},\hspace{0.5cm}\ket{\psi^c(t)}\rightarrow\ket{\psi^c(0)}e^{-iE_{\nu^c}t}.\nonumber
\label{5.4}
\end{equation}
In the ultrarelativistic limit $|\textbf{p}|>>m$ (in natural units, $\hbar=c=1$), the survival probability of the initial state $\ket{\nu_1}$ at any time $t$ can be written as
\begin{equation}
    P_s(t)=\left|\left\langle\left. \nu_1(t) \right| \nu_1(0) \right\rangle \right|^2=1-\sin^2(2\theta_\nu) \sin^2\left(\delta t\right),
   \label{5.5}
\end{equation}
where the term
\begin{equation}
    \delta=\frac{1}{2}(E_\nu-E_{\nu^c})\approx (B^0-|\textbf{B}|)
    \label{5.5a}
\end{equation}
represents half of the energy difference between neutrino and antineutrino. Subsequently, the oscillation probability of the initial state $\nu_1$ at any time $t$ is $P_O(t)=\left|\left\langle\left. \nu_2(t) \right| \nu_1(0) \right\rangle \right|^2=1-P_S(t)$. Thus, $P_S(t)\neq 0\Longrightarrow P_O(t)\neq 0$. 

An important point to note here is that in arriving at the energy expressions given in Eq.\,(\ref{5.3}), we have assumed a local plane wave solution with a constant $B^d$. This approximation is justified because we are working in the WKBJ approximation regime. The WKBJ condition is given by \cite{Sakurai:2011zz} 
\begin{equation}
    \lambda=\frac{\hslash}{p}<<\frac{pc}{\left|\frac{dB^d}{dr}\right|}
    \label{wkb}
\end{equation}
From Figs.\,\ref{Fig2} and \ref{Fig3}, it is evident that Eq.\,(\ref{wkb}) is satisfied with our ultrarelativistic assumption. This implies that the wavelength of the neutrinos is small enough compared to the length scale over which the spin-curvature coupling varies significantly. Thus, the coupling can be considered constant over small patches of the spacetime. This enables us to consider plane wave neutrino wave functions locally, which can then be stitched together to obtain a global solution. We use this framework to determine the global spatial evolution of various quantities related to neutrino oscillations.

Using Eq.\,(\ref{5.5a}), in the top panel of Fig.\,\ref{Fig6}, we illustrate the variation of $|\delta|$ with $r$ for different values of $\theta$, with $a=0.998$ and $m=0.0497666\rm\,{eV}$. We observe that the energy difference between neutrino and antineutrino ($|\delta|$) is large at a small radial distance $r$ from the PBH. It is because, at small radial distances from the PBH, the strengths of $B^0$ and $|\textbf{B}|$ are very high, though unequal (see the top panels of Figs.\,{\ref{Fig2}} and {\ref{Fig3}}), leading to a large energy difference. In contrast, at large radial distances from the PBH, $B^0$ and $|\textbf{B}|$ decrease, causing $|\delta|$ to decrease as well. Since the strengths of $B^0$ and $|\textbf{B}|$ are largest for $\theta=\pi/3$ as compared to other $\theta$ throughout the range of $r$ (see the blue dashed line in the top panels of Figs.\,{\ref{Fig2}} and {\ref{Fig3}}), $|\delta|$ remains largest at this value of $\theta$ (green dashed line), compared to other values of $\theta$, as illustrated in the top panel of Fig.\,\ref{Fig6}. In the bottom panel of Fig.\,\ref{Fig6}, we show the variation of $|\delta|$ with $r$ for different values $a$ with $\theta=\pi/4$ and $m=0.0497666\rm\,{eV}$. Since at $a=0.998$, $B^0$ and $|\textbf{B}|$ remain highest among the $a$ considered throughout the range of $r$  (see the magenta dotted line in the bottom panels of Figs.\,{\ref{Fig2}} and {\ref{Fig3}}), we observe that $|\delta|$ remains high across the entire range of $r$ for high values of $a$ (see the black solid line in the bottom panel of Fig.\,\ref{Fig6}), compared to low values of $a$.

Furthermore, we consider an observer at infinity and deduce the relation between neutrino propagation time $t$ to the radial distance $r$. To do so, we consider neutrinos moving along radial trajectories. We find the relation between $t$ and $r$ from the line element given by Eq.\,(\ref{1}) by setting $ds^2=0$ with $d\phi=0$ and $d\theta=0$.  Consequently, we obtain
\begin{equation}
    t=\int^r_{r_1> r_{+}}f(r,\theta,a)dr,
    \label{5.6}
\end{equation}
where,
\begin{equation*}
    f(r,\theta,a)=\frac{r^2+2r+a^2\cos^2\theta}{r^2-2r+a^2\cos^2\theta},\nonumber
\end{equation*}
    and $r_+$ is the outer event horizon of the spinning PBH given by Eq.\,({\ref{2}}).
    Subsequently, using Eq.\,(\ref{5.6}) in Eq.\,(\ref{5.5}), the survival probability of the initial state $\ket{\nu_1}$ at any time $t$ can be expressed as a function of radial distance $r$ as
    \begin{equation}
P_S(r)=1-\sin^2(2\theta_\nu) \sin^2\left[\delta\int^r_{r_1> r_{+}}f(r,\theta,a)dr\right].
\label{5.7}
    \end{equation}
In the top panel of Fig.\,{\ref{Fig7}}, we illustrate the survival probability $P_S$ of the initial state $\ket{\nu_1}$ as a function of the radial distance $r$ for different values of $\theta$, with $a=0.998$ and $m=0.0497666\rm\,{eV}$. As shown in the top panel of Fig.\,\ref{Fig2}, at $\theta=\pi/2$ (red dot-dashed line), the contribution of $B^0$ is zero, resulting in the saturation of $\theta_\nu$ at $\pi/4$, as evidenced by the red dot-dashed line in the top panel of Fig.\,\ref{Fig5}. This leads to a change in $P_S$  (see the red solid line in the top panel of Fig.\,{\ref{Fig7}}) with the maximum mixing angle ($\theta_\nu=\pi/4$). The behavior of $P_S$ for $\theta=\pi/2$ is thus governed by $\delta$. The high value of $|\delta|$ (see red dot-dashed line in Fig.\,{\ref{Fig6}}) near the PBH ($r\approx3r_g$), leads to the incomplete transition of $P_S$ between 1 and 0.2 as depicted by the red solid line in the top panel of Fig.\,\ref{Fig7}. At larger distances from the PBH, the magnitude of $\delta$ decreases, leading to complete oscillations between 1 and 0. 
However, for other values of $\theta$ (say $\pi/3$ , $\pi/4$, and $10\pi/18$) near the PBH ($r\leq 6r_g$), the mixing angle ($\theta_\nu$) varies approximately between 0 and $\pi/6$ (see the top panel of Fig.\,\ref{Fig5}) and $|\delta|$ (see the top panel of Fig.\,\ref{Fig6}) is large, causing the amplitude of $P_S$ to change gradually between 1 and 0 in the top panel of Fig.\,\ref{Fig7}. These conditions suppress the full change in $P_S$ between 1 and 0 near the PBH ($r\leq 6r_g$). At larger radial distances from the PBH, $\theta_\nu$ saturates to the maximum mixing angle of $\pi/4$ (see the top panel of Fig.\,{\ref{Fig5}}) for all values of $\theta$. Additionally, $|\delta|$ decreases at large $r$ (see the top panel of Fig.\,{\ref{Fig6}}) and becomes approximately the same for all $\theta$. It is important to note that if $\delta\rightarrow 0$, the oscillation length ($\pi$/$|\delta|$) will increase. These conditions cause the oscillatory terms in Eq.\,({\ref{5.7}}) to vanish, i.e., $\sin^2\left[\delta\int^r_{r_1> r_{+}}f(r,\theta, a)dr\right]\rightarrow 0$ at large $r$. Consequently, in the top panel of Fig.\,{\ref{Fig7}}, $P_s(r)$ (see, $P_S$ in Eq.\,({\ref{5.7}})) tends to 1 and coincides for all values of $\theta$ as the radial distance $r$ increases to $r\approx 100 r_g$. Hence, up to the radial distance $r\approx100 r_g$, oscillations in the neutrino-antineutrino system induced by gravity are highly dependent on the angle of the position vector of the neutrino spinor with respect to the spin axis of the PBH. However, at larger distances from the PBH ($r>100 r_g$), the strength of the four-vector gravitational potential diminishes, which causes the survival probability $P_S$ to become independent of the different values of $\theta$. 

The bottom panel of Fig.\,{\ref{Fig7}} shows $P_s$ as a function of $r$ for different values of $a$, with $\theta=\pi/4$ and $m=0.0497666\rm\,{eV}$. As shown in the bottom panel of Fig.\,\ref{Fig5}, the change in the behavior of the mixing angle $\theta_\nu$ is sharper for a low value $a=0.1$ (red dot-dashed line) and more shallow for a high value $a=0.998$ (magenta dotted line).  Furthermore, the strength of $|\delta|$ (see the bottom panel of Fig.\,{\ref{Fig6}}) is higher for larger values of $a$ throughout the range of $r$. 
It is also evident from the bottom panel of Fig.\,{\ref{Fig7}} that the oscillation length increases with decreasing $a$. $P_S$ of the initial neutrino state $\ket{\nu_1}$ at $a=0.998$ (black solid line) oscillates back to 1 more frequently very near the PBH ($\leq$ 6 $r_g$), owing to the short oscillation length caused by the high strength of $|\delta|$. However, far from the PBH, $P_S$ oscillates less frequently back to 1 over a large radial distance $r$, exhibiting a longer oscillation length due to the low strength of $|\delta|$.

These results indicate that at $\theta=\pi/4$ and for a large value of $a$ (black solid line), the neutrino-antineutrino system is less prone to oscillations near the PBH but more prone to oscillations over a large range of radial distances $r$ from the PBH. Additionally, in the bottom panel of Fig.\,\ref{Fig5}, at $a=0.1$ (red dot-dashed line), $\theta_\nu$ saturates to $\pi/4$ at smaller $r$ than that for a high value $a=0.998$ (magenta dotted line). Moreover, the strength of $|\delta|$ in the bottom panel of Fig.\,\ref{Fig6} is high near the PBH and low far away from the PBH at $a=0.1$ (red dot-dashed line). As a result, in the bottom panel of Fig.\,\ref{Fig7}, $P_S$ (red dot-dashed line) initially varies between 1 and 0.5 near the PBH ($r\approx 3.5r_g$) but undergoes a complete transition between 1 and 0 farther from the PBH. Since $|\delta|$ is lower across the entire range of $r$ at $a=0.1$ (see the red dot-dashed line in the bottom panel of Fig.\,\ref{Fig6}), compared to other values of $a$, the corresponding oscillation length 
remains longer throughout the range of $r$. However, at large radial distances from the PBH ($r> 100 r_g$), since $B^0$ and $|\textbf{B}|$ diminish (see the bottom panel of Fig.\,\ref{Fig2} and Fig.\,\ref{Fig3}), the strength of $|\delta|$ tends to 0 for all values of $a$ (see the bottom panel of Fig.\,\ref{Fig6}), which causes the oscillatory terms in Eq.\,({\ref{5.7}}) to vanish i.e., $\sin^2\left[\delta\int^r_{r_1> r_{+}}f(r,\theta, a)dr\right]\rightarrow 0$. Consequently, in the bottom panel of Fig.\,{\ref{Fig7}}, $P_S(r)$ [see Eq.\,({\ref{5.7}})] tends to 1 and becomes independent of $a$ at $r> 100 r_g$. Therefore, oscillations in the neutrino-antineutrino system induced by gravity are highly dependent on the strength of $a$ only within the short range of radial distance $r$ ($r<100r_g$) from the PBH.
 
We observe that the primary cause of neutrino-antineutrino transitions is the nonzero magnitude of the four-vector gravitational potential ($B^0$ and $|\textbf{B}|$). The four-vector gravitational potential increases near the PBH, resulting in an increase in the space-time coupling of neutrinos. This strong coupling suppresses the oscillation of the neutrino-antineutrino system. It is evident from the top and bottom panels of Fig.\,{\ref{Fig7}} that $P_s$ reaches its maximum value of 1 very close to the PBH ($r\approx 3r_g$) for all the trajectories and spin parameters. Additionally, in the bottom panel of Fig.\,{\ref{Fig7}}, we observe that the neutrino oscillation probability ($1-P_s$) increases at smaller radii ($3r_g<r<4r_g$) as the PBH spin decreases. This could be attributed to the increased gravitational field strength with increase in $a$, leading to an enhanced coupling between the neutrinos and the spacetime curvature, making the system less oscillatory at a given radius. 

The oscillations in the neutrino-antineutrino system in curved spacetime discussed here represent a simplified toy model. A more realistic approach involves considering the two-flavor oscillations of the neutrino-antineutrino system in curved spacetime, which we discuss next.

\section{Two-flavor oscillations of neutrino-antineutrino system}
\label{sec7}
Following the treatment provided in Sec.\,\ref{sec5} and using the earlier method (see Sec.\,IV of Ref.\,\cite{Sinha:2007uh}), in the background gravitational field, the Euler-Lagrangian equation in four spinor component form for the two flavor case is
\begin{equation}
    i\gamma^0\gamma^c \mathcal{\partial}_c\begin{pmatrix}
        \psi^c_e\\
        \psi^c_\mu\\
        \psi_e\\
        \psi_\mu\\
    \end{pmatrix}+\mathcal{M}_4\begin{pmatrix}
        \psi^c_e\\
        \psi^c_\mu\\
        \psi_e\\
        \psi_\mu\\
    \end{pmatrix}=0,
    \label{7.1}
    \end{equation}
where $\psi_e$, $\psi_\mu$, $\psi^c_e$, and $\psi^c_\mu$ are corresponding flavor spinors for electron neutrino, muon neutrino, electron antineutrino, and muon antineutrino, respectively. Thus, in the case of two flavor scenario, the most general flavor mixing mass matrix in the presence of gravity can be written as 
\begin{equation}
    \mathcal{M}_4=\begin{pmatrix}
        -B^0 \mathbf{I}- B^i\sigma_i & -\mathbf{M}\\
        -\mathbf{M} &  B^0 \mathbf{I}- B^i\sigma_i\\
    \end{pmatrix},
    \label{7.2}
\end{equation}
where $\mathbf{I}$ is the $2\times2$ unit matrix and 
\begin{equation}
    \mathbf{M}=\begin{pmatrix}
        m_e & m_{e\mu}\\
        m_{e\mu} & m_\mu\\
    \end{pmatrix}\equiv \mathbf{U(\theta_v)} . \text{diag} (m_1,m_2). \mathbf{U^\dagger(\theta_v)}.
    \label{7.3}
\end{equation}
Here, the neutrino masses and the mixing matrix in vacuum are denoted by $m_{1,2}$ and $\mathbf{U(\theta_v)}$ (where $\theta_v$ is the mixing angle in vacuum), respectively. The Majorana masses for the electron and muon neutrinos are $m_{e}$ and $m_{\mu}$, respectively, and $m_{e\mu}$ is the Majorana mixing mass. The mass matrix $\mathcal{M}_4$, given by Eq.\,(\ref{7.2}) can be diagonalized by a $4\times4$ unitary matrix $\mathbf{T}$. Consequently, at time $t=0$, each flavor state can be written in the superposition of mass eigenstates as \cite{Sinha:2007uh}
\begin{equation}
    \begin{pmatrix}
       \ket{\psi_e^c}\\
        \ket{\psi_\mu^c}\\
        \ket{\psi_e}\\
        \ket{\psi_\mu}\\
    \end{pmatrix}=\mathbf{T}\begin{pmatrix}
        \ket{\chi_1}\\
        \ket{\chi_2}\\
        \ket{\chi_3}\\
        \ket{\chi_4}\\
    \end{pmatrix},
    \label{7.4}
\end{equation}
where the mixing matrix $\mathbf{T}$ can be represented as
\begin{small}
\begin{equation}
    \mathbf{T}=\begin{pmatrix}
        \cos\theta_e \cos\phi_1 & -\cos\theta_e \sin\phi_1 & -\sin\theta_e \cos\phi_2 & \sin\theta_e \sin\phi_2\\
        \cos\theta_\mu \sin\phi_1 & \cos\theta_\mu \cos\phi_1 & -\sin\theta_\mu \sin\phi_2 & -\sin\theta_\mu \cos\phi_2\\
        \sin\theta_e \cos\phi_1 & -\sin\theta_e \sin\phi_1 & \cos\theta_e \cos\phi_2 & -\cos\theta_e \sin\phi_2\\
        \sin\theta_\mu \sin\phi_1 & \sin\theta_\mu \cos\phi_1 & \cos\theta_\mu \sin\phi_2 & \cos\theta_\mu \cos\phi_2\\
    \end{pmatrix}.
    \label{7.5}
\end{equation}
\end{small}
The time evolution of mass eigenstates is given by
\begin{equation}
    \begin{pmatrix}
       \ket{ \chi_1(t)}\\
        \ket{\chi_2(t)}\\
        \ket{\chi_3(t)}\\
        \ket{\chi_4(t)}\\
    \end{pmatrix}=\text{diag}(e^{-iE_1 t},e^{-iE_2 t},e^{-iE_3 t},e^{-iE_4 t})\begin{pmatrix}
        \ket{\chi_1}\\
        \ket{\chi_2}\\
        \ket{\chi_3}\\
        \ket{\chi_4}\\
    \end{pmatrix},
    \label{7.6}
\end{equation}
 where $\ket{\chi_j}$ ($j=1,2,3,4$) is the mass eigenstate at time $t=0$. Then using Eqs.\,(\ref{7.4}), (\ref{7.5}), and (\ref{7.6}), each flavor state at some later time $t$ can be expressed as
\begin{align}
     & \begin{pmatrix}
       \ket{\psi_e^c(t)}\\
        \ket{\psi_\mu^c(t)}\\
        \ket{\psi_e(t)}\\
        \ket{\psi_\mu(t)}\\
    \end{pmatrix}=\mathbf{T}\begin{pmatrix}
       \ket{\chi_1(t)}\\
        \ket{\chi_2(t)}\\
        \ket{\chi_3(t)}\\
        \ket{\chi_4(t)}\\
    \end{pmatrix}\nonumber\\
    &=\mathbf{T}\,\text{diag}(e^{-iE_1 t},e^{-iE_2 t},e^{-iE_3 t},e^{-iE_4 t}) \begin{pmatrix}
       \ket{\chi_1}\\
        \ket{\chi_2}\\
        \ket{\chi_3}\\
        \ket{\chi_4}\\
    \end{pmatrix}.
    \label{7.7}
\end{align}
Since, $\mathbf{T} \mathbf{T}^\dagger=\textbf{I}\Rightarrow \mathbf{T}^\dagger=\mathbf{T}^{-1}$. Consequently, by inverting the mixing matrix $\mathbf{T}$ and using Eq.\,(\ref{7.4}), we can write
\begin{equation}
   \begin{pmatrix}
        \ket{\chi_1}\\
        \ket{\chi_2}\\
        \ket{\chi_3}\\
        \ket{\chi_4}\\
    \end{pmatrix} =\mathbf{T}^{-1} \begin{pmatrix}
       \ket{\psi_e^c}\\
        \ket{\psi_\mu^c}\\
        \ket{\psi_e}\\
        \ket{\psi_\mu}\\
    \end{pmatrix}.
    \label{7.4a}
\end{equation}
Substituting Eq.\,(\ref{7.4a}) into Eq.\,(\ref{7.7}), the evolving flavor states $\ket{\psi_e^c(t)}$, $\ket{\psi_\mu^c(t)}$, $\ket{\psi_e(t)}$ and $\ket{\psi_\mu(t)}$ can also be projected to flavor basis in the form 
\begin{equation}
     \begin{pmatrix}
       \ket{\psi_e^c(t)}\\
        \ket{\psi_\mu^c(t)}\\
        \ket{\psi_e(t)}\\
        \ket{\psi_\mu(t)}\\
    \end{pmatrix}=\mathbf{T}\,\text{diag}(e^{-iE_1 t},e^{-iE_2 t},e^{-iE_3 t},e^{-iE_4 t})\, \mathbf{T}^{-1}\begin{pmatrix}
       \ket{\psi_e^c}\\
        \ket{\psi_\mu^c}\\
        \ket{\psi_e}\\
        \ket{\psi_\mu}\\
    \end{pmatrix},
    \label{7.8}
\end{equation}
\begin{table}[h!]
\centering
\begin{tabular}{ |p{0.3\linewidth}|p{0.5\linewidth}|}
 \hline
  Parameters & Best fit$\pm 1\sigma$  \\ 
\hline
$\theta_{12}$ &  $(33.66\pm 0.72)^\circ$  \\
 \hline
 $\Delta m^2_{21}$ &$(7.53\pm0.18)\times 10^{-5}\rm\,{eV^2}$\\
 \hline
 $\Delta m^2_{32}$  & $(-2.529 \pm 0.029)\times 10^{-3}\rm\,{eV^2}$ \\
 \hline
\end{tabular}
\caption{\justifying{The values of the neutrino mixing parameters for IH that we considered in our analysis, taken from PDG \cite{ParticleDataGroup:2024cfk} along with their corresponding $1\sigma$ errors ($90\%$ CL).}}
\label{tab1}
\end{table}where $\ket{\psi_e^c}$, $\ket{\psi_\mu^c}$, $\ket{\psi_e}$ and $\ket{\psi_\mu}$ are flavor states at time $t=0$. Here,
\begin{small}
\begin{equation}
    \mathbf{T}\,.\text{diag}(e^{-iE_1 t},e^{-iE_2 t},e^{-iE_3 t},e^{-iE_4 t}).\, \mathbf{T}^{-1}=\begin{pmatrix}
        T_{e^c e^c}(t) & T_{e^c\mu^c}(t) & T_{e^c e}(t) & T_{e^c\mu}(t)\\
        T_{\mu^ce^c}(t) & T_{\mu^c\mu^c}(t) & T_{\mu^ce}(t) & T_{\mu^c\mu}(t)\\
        T_{ee^c}(t) & T_{e\mu^c}(t) & T_{ee}(t) & T_{e\mu}(t)\\
        T_{\mu e^c}(t) & T_{\mu \mu^c}(t) & T_{\mu e}(t) & T_{\mu\mu}(t)\\
    \end{pmatrix}.
    \label{7.8a}
\end{equation}
\end{small}Thus, substituting Eq.\,(\ref{7.8a}) into Eq.\,(\ref{7.8}), the time-evolved electron flavor neutrino state in the superposition of flavor basis can be obtained as
\begin{equation}
   \ket{\psi_e(t)}=T_{ee^c}(t)\ket{\psi^c_e} +T_{e\mu^c}(t)\ket{\psi^c_\mu}
   + T_{ee}(t) \ket{\psi_e}+T_{e\mu}(t) \ket{\psi_\mu},
   \label{7.9}
\end{equation}
where $|T_{ee^c}(t)|^2+|T_{e\mu^c}(t)|^2+|T_{ee}(t)|^2+|T_{e\mu}(t)|^2=1$ and the coefficients are 
\begin{align}
    T_{ee^c}(t)&=\cos\theta_e\sin\theta_e(e^{-iE_1 t}\cos^2\phi_1-e^{-iE_3 t}\cos^2\phi_2\nonumber\\
    &+e^{-iE_2t}\sin^2\phi_1-e^{-iE_4t}\sin^2\phi_2),\nonumber\\
    T_{e\mu^c}(t)&=\cos\phi_1 \cos\theta_\mu\sin\phi_1\sin\theta_e(e^{-iE_1t}-e^{-iE_2t})\nonumber\\
    & +\cos\phi_2\cos\theta_e\sin\phi_2\sin\theta_\mu(e^{-iE_4t}-e^{-iE_3 t}),\nonumber\\
   T_{ee}(t)&=\cos^2\theta_e(e^{-iE_3 t}\cos^2\phi_2+e^{-iE_4 t}\sin^2\phi_2)\nonumber\\
    &+\sin^2\theta_e(e^{-iE_1t}\cos^2\phi_1+e^{-iE_2t}\sin^2\phi_1),\nonumber\\
T_{e\mu}(t)&=\cos\phi_1\sin\theta_\mu\sin\phi_1\sin\theta_e(e^{-iE_1t}-e^{-iE_2t})\nonumber\\
&-\cos\phi_2\cos\theta_e\sin\phi_2\cos\theta_\mu(e^{-iE_4t}-e^{-iE_3t}).
\label{7.10}
\end{align}
The mixing angles for neutrino-antineutrino ($\theta_{e,\mu}$) and electron-muon neutrino mixing ($\phi_{1,2}$) are related to the masses and the gravitational potential, respectively, by
\begin{equation}
    \theta_{e,\mu}=\tan^{-1}\left[\frac{m_{e,\mu}}{B^0+\sqrt{(B^0)^2+m^2_{e,\mu}}}\right],
    \label{7.11}
\end{equation}
\begin{equation}
    \phi_{1,2}=\tan^{-1}\left[\frac{\mp 2 m_{e\mu}}{m_{e(1,2)}-m_{\mu(1,2)}+\sqrt{(m_{e(1,2)}-m_{\mu(1,2)})^2+4m^2_{e\mu}}}\right].
    \label{7.12}
\end{equation}

\begin{figure}[!htbp]
   \centering
    \begin{subfigure}{0.99\columnwidth}
       \centering
       \caption{}
        \label{Fig8a}
       \includegraphics[width=\textwidth]{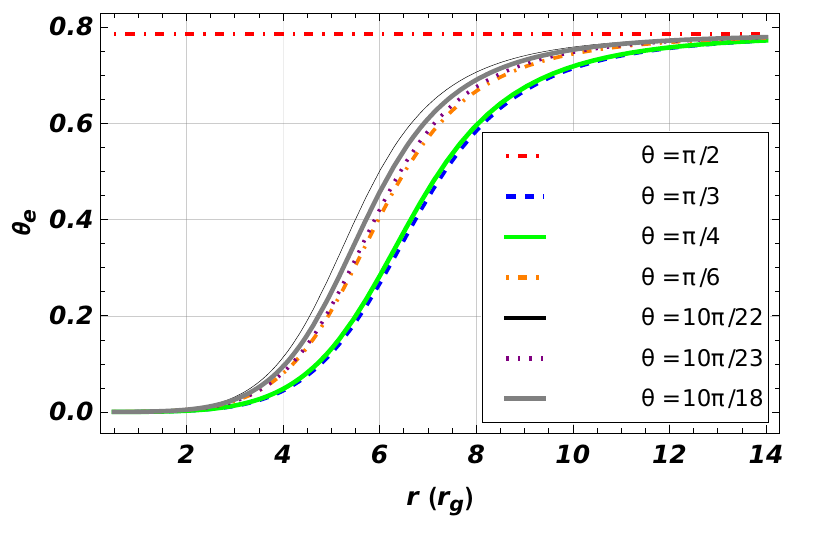}
    \end{subfigure}
   \hfill
   \begin{subfigure}{0.99\columnwidth}
        \centering
        \caption{}
        \label{Fig8b}
        \includegraphics[width=\textwidth]{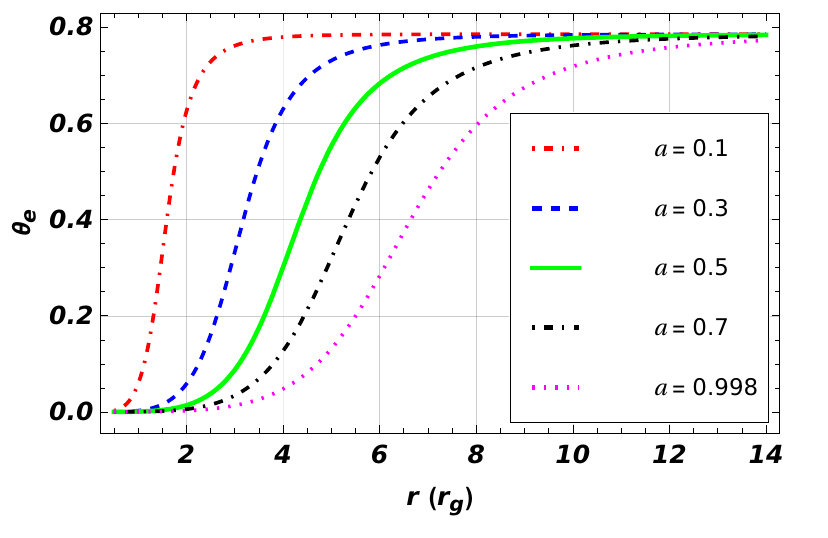}
    \end{subfigure}
   \caption{\justifying{In the two-flavor scenario, the mixing angle $\theta_e$ as a function of the radial distance $r$: (a) in the top panel, for different values of angle $\theta$, with $a=0.998$; and (b) in the bottom panel, for different values of $a$, with $\theta=\pi/4$, respectively.}}
    \label{Fig8}
\end{figure}

The masses corresponding to the mass eigenstates are given as \cite{Sinha:2007uh}
\begin{eqnarray}
    {M_{1,2}=\frac{1}{2}[(m_{e1}+m_{\mu 1})\pm\sqrt{(m_{e1}-m_{\mu 1})^2+4m^2_{e\mu}}],}&\nonumber\\
    {M_{3,4}=\frac{1}{2}[(m_{e2}+m_{\mu 2})\pm\sqrt{(m_{e2}-m_{\mu 2})^2+4m^2_{e\mu}}]},&
    \label{7.13}
\end{eqnarray}
with
\begin{small}
\begin{align}
    m_{(e,\mu)1}&=-\sqrt{(B^0)^2+|\textbf{B}|^2+m^2_{e,\mu}+2\sqrt{|\textbf{B}|^2((B^0)^2+m^2_{e,\mu})}},\nonumber\\
   m_{(e,\mu)2} &=\sqrt{(B^0)^2+|\textbf{B}|^2+m^2_{e,\mu}+2\sqrt{|\textbf{B}|^2((B^0)^2+m^2_{e,\mu})}},
    \label{7.14}
\end{align}
\end{small}are two eigenvalues obtained from Eq.\,(\ref{4.6}).
The energies corresponding to mass eigenstates with momentum $\vec{p}$ are 
\begin{align}
E_1&=\sqrt{(\textbf{p}+\textbf{B})^2+M_1^2}-B^0,\hspace{0.3cm}E_2=\sqrt{(\textbf{p}+\textbf{B})^2+M_2^2}-B^0,\nonumber\\
   E_3&=\sqrt{(\textbf{p}-\textbf{B})^2+M_3^2}+B^0,\hspace{0.3cm}
   E_4=\sqrt{(\textbf{p}-\textbf{B})^2+M_4^2}+B^0.
   \label{7.15}
\end{align}
Moreover, there are two equivalent orderings for the spectrum of neutrino masses. The first one is a spectrum with normal ordering (NO) with $m_1<m_2<m_3$, and the second one is a spectrum with inverted ordering (IO) with $m_3<m_1<m_2$. The Particle Data Group (PDG) shows a hierarchy between the mass splittings \cite{ParticleDataGroup:2024cfk}, $\Delta m^2_{21}<<|\Delta m^2_{31}|\simeq|\Delta m^2_{32}|$ with $\Delta m^2_{ij}\equiv m^2_i-m^2_j$. Note that $\Delta m^2_{32}>0$ for NO and $\Delta m^2_{32}<0$ for IO.

\begin{figure}[!htbp]
   \centering
    \begin{subfigure}{0.99\columnwidth}
       \centering
       \caption{}
        \label{Fig10a}
       \includegraphics[width=\textwidth]{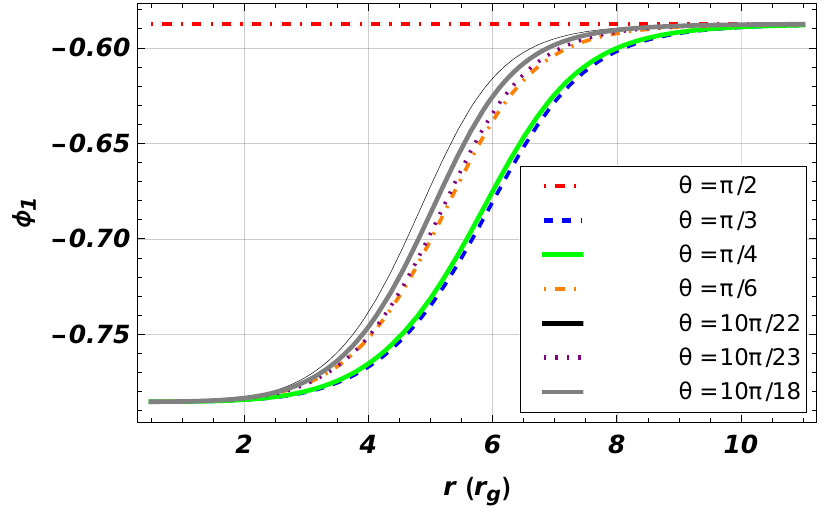}
    \end{subfigure}
   \hfill
   \begin{subfigure}{0.99\columnwidth}
        \centering
        \caption{}
        \label{Fig10b}
        \includegraphics[width=\textwidth]{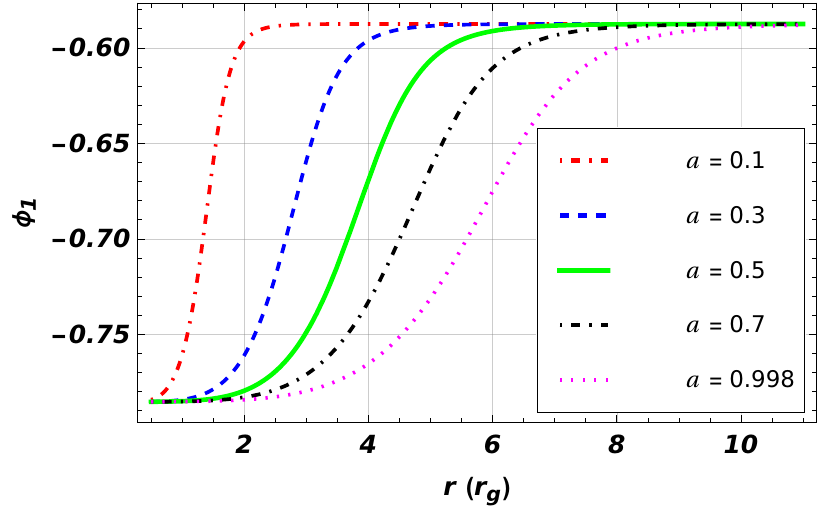}
    \end{subfigure}
   \caption{\justifying{In the two-flavor scenario, the mixing angle $\phi_1$ as a function of the radial distance $r$: (a) in the top panel, for different values of angle $\theta$,
   with $a=0.998$; and (b) in the bottom panel, for different values of $a$, with $\theta=\pi/4$, respectively.}}
    \label{Fig10}
\end{figure}

Based on the value of the lightest neutrino mass, the neutrino mass spectrum can be further classified in normal hierarchical (NH) spectrum: $m_1<<m_2<m_3$, and inverted hierarchical (IH) spectrum $m_3<<m_1<m_2$. In the IH spectrum, the neutrino mixing parameters $\Delta m^2_{ij}$ and $\theta_{12}$, along with their $1\sigma$ best-fit data, is presented in Table\,\ref{tab1}. Subsequently, we get
$m_1\simeq\sqrt{|\Delta m^2_{32}+\Delta m^2_{21}|}\sim 0.0495\pm 3.1 \times 10^{-4}\rm\, {eV}$ and $m_2\simeq\sqrt{|\Delta m^2_{32}|}\sim 0.05\pm 2.867\times 10^{-4}\rm\,{eV}$ \cite{ParticleDataGroup:2024cfk}. Using these $m_1$, $m_2$, and the vacuum mixing angle $\theta_v\equiv\theta_{12}$ in Eq.\,(\ref{7.3}), we calculate the value of the Majorana mass\footnote{Throughout the paper, we use the calculated values of the Majorana mass as required.} as $m_e=0.0497666\pm 3.115\times 10^{-4}\rm\,{eV}$, $m_\mu=0.0500574\pm 3.02\times 10^{-4}\rm\,{eV}$, and $m_{e\mu}=0.000347999\pm 1.01\times 10^{-5}\rm\,{eV}$. 
Thus, by substituting these Majorana masses into Eqs.\,(\ref{7.11}) and (\ref{7.12}), we calculate $\theta_e$, $\theta_\mu$, $\phi_1$, and $\phi_2$.

The top and bottom panels of Fig.\,\ref{Fig8} illustrate the variation of the mixing angle $\theta_e$ with $r$, for different values of $\theta$ with fixed $a=0.998$ (top panel) and different values of $a$ with fixed $\theta=\pi/4$ (bottom panel), respectively. The mixing angle $\theta_\mu$ follows the same nature as $\theta_e$. The behavior of these mixing angles is observed to be similar to that of the neutrino-antineutrino mixing angle, as illustrated in Fig.\,\ref{Fig5}.

Similarly, the variation in the mixing angle $\phi_1$ with $r$ for different values of $\theta$ and $a$ are shown, respectively, in the top and bottom panels of Fig.\,\ref{Fig10}. We also find that the behavior of $\phi_2$ for different values of $a$ and $\theta$ resembles that of $\phi_1$ but with variation between 0.80 and 0.95.

\begin{figure}[!htbp]
   \centering
    \begin{subfigure}{0.99\columnwidth}
       \centering
       \caption{}
        \label{Fig13a}
       \includegraphics[width=\textwidth]{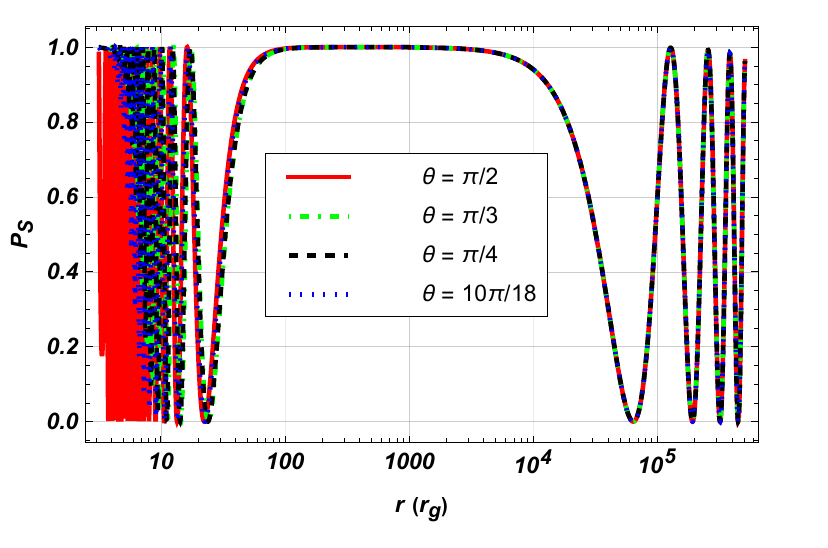}
    \end{subfigure}
   \hfill
   \begin{subfigure}{0.99\columnwidth}
        \centering
        \caption{}
        \label{Fig13b}
        \includegraphics[width=\textwidth]{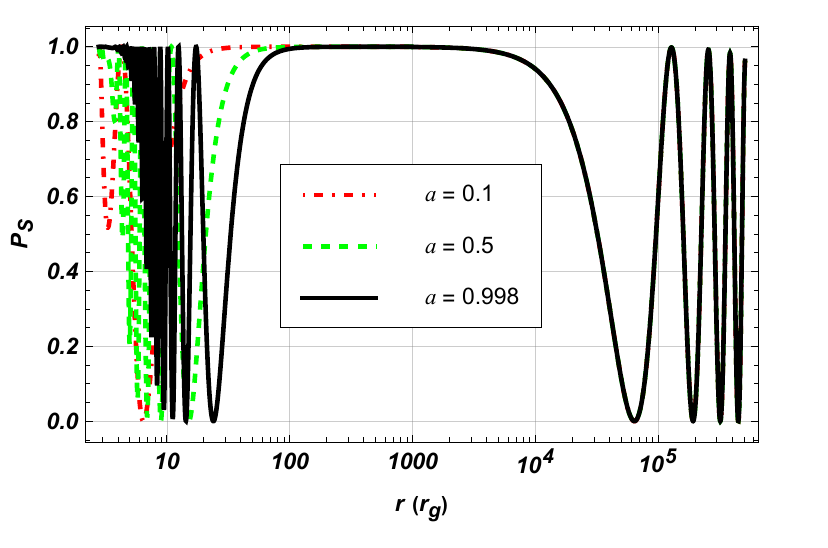}
    \end{subfigure}
   \caption{\justifying{In the two-flavor scenario, the survival probability $P_S$ as a function of the radial distance $r$: (a) in the top panel, for different values of angle $\theta$,
   with $a=0.998$; and (b) in the bottom panel, for different values of $a$,
   with $\theta=\pi/4$.}}
    \label{Fig13}
\end{figure}

Furthermore, in the ultrarelativistic limit, Eq.\,(\ref{7.15}), in the first-order approximation, reduces to 
\begin{align}
    E_1&=E_2=|\textbf{p}|+|\textbf{B}|-B^0,\nonumber\\
   E_3&=E_4=|\textbf{p}|-|\textbf{B}|+B^0.
   \label{7.15a}
\end{align}
  When an observer is at infinity, we find the survival probability of the initial electron flavor neutrino state $\ket{\psi_e}$ as a function of radial distance $r$ using the coefficient $T_{ee}(t)$ given by Eq.\,(\ref{7.10}) and following Eqs.\,(\ref{7.11})-(\ref{7.15a}) with Eq.\,(\ref{5.6}) as
\begin{equation}
    P_s(r)=|T_{ee}(r)|^2.
    \label{7.16}
\end{equation}
Using Eq.\,(\ref{7.16}), in the top panel of Fig.\,{\ref{Fig13}}, we illustrate the survival probability ($P_S$) of the initial electron flavor neutrino state $\ket{\psi_e}$ as a function of the radial distance $r$, for different values of $\theta$, with $a=0.998$. At a small radial distance $r$ from the PBH, $P_S$ changes less frequently between 1 and 0 for all values of $\theta$ (except for $\theta=\pi/2$). This is because the strength of $B^0$ and $|\textbf{B}|$ are high for all $\theta$ in this regime (see the top panel Fig.\,{\ref{Fig2}} and Fig.\,{\ref{Fig3}}), but the mixing angles such as $\theta_e$, $\theta_\mu$, $\phi_1$ and $\phi_2$ are relatively small (see the top panel of Figs.\,{\ref{Fig8}} and {\ref{Fig10}}) causing $P_S$ to change less frequently. It is important to note that, particularly at $\theta=\pi/2$, the magnitudes of $B^0$, $B^1$, and $B^3$ are zero, while $B^2$ is nonzero. Therefore, due to $B^2$, $P_S$ (red solid line) changes between 1 and 0 at $\theta=\pi/2$. In contrast, at large radial distances $r$ from the PBH, the strengths of $B^0$ and $|\textbf{B}|$ decrease for all values of $\theta$ (see the top panel Fig.\,{\ref{Fig2}} and Fig.\,{\ref{Fig3}}). As a result, the mixing angles, such as $\theta_e$, $\theta_\mu$, $\phi_1$ and $\phi_2$ begin to vary more with $r$ and after covering some distance, they saturate (see the top panel of Figs.\,{\ref{Fig8}} and {\ref{Fig10}}). This leads to a variation in the oscillation behavior of $P_S$ at each point of $r$, causing $P_s$ to change more frequently between 1 and 0 for all values of $\theta$. 

Similarly, in the bottom panel of Fig.\,{\ref{Fig13}}, we depict the survival probability ($P_S$) of the initial electron flavor neutrino state $\ket{\psi_e}$ as a function of the radial distance $r$ from the PBH for different values of $a$ with a fixed $\theta=\pi/4$. Near the PBH, a large $a$ corresponds to high strengths of $B^0$ and $|\textbf{B}|$ (see the magenta dotted line in the bottom panels of Figs.\,{\ref{Fig2}} and {\ref{Fig3}}) and vice-versa. Consequently, $P_s$ changes between 1 and 0 less frequently near the PBH and more frequently far away from the PBH at $a=0.998$ (black solid line) compared to the other values of $a$. Thus, $P_S$ becomes 1 for all $a$ and $\theta$ due to the strong gravitational field near the PBH. However, we observe that $P_S$ becomes independent of both $\theta$ and $a$ (see the top and bottom panels of Fig.\,{\ref{Fig13}}) at $r\approx 100 r_g$.
It is because, near $r\approx 100 r_g$, the strength of $B^0$ and $|\textbf{B}|$ diminishes. 

The question remains how the different values of $\theta$ and $a$ affect $P_S$ when observers are at extremely long distances from the PBH, say $r>>100 r_g$. To address this question, in the ultrarelativistic regime, the first-order approximation of the energies given by Eq.\,(\ref{7.15a}) is insufficient. This is due to $B^d$ being significantly smaller in comparison to the mass eigenvalues ($M_1, M_2, M_3, \text{and}~M_4$).
Therefore, the second-order approximation of the energies in the survival probability ($P_S$) for the initial state $\ket{\psi_e}$ must be included. The second-order approximation has four different energies ($E_1\neq E_2\neq E_3 \neq E_4$), corresponding to four different mass eigenstates. This approximation can be derived from Eq.\,(\ref{7.15}) as 
\begin{align}
E_1&=|\textbf{p}|+|\textbf{B}|+\frac{|\textbf{B}|^2+M_1^2}{2|\textbf{p}|}-B^0,\nonumber\\
E_2&=|\textbf{p}|+|\textbf{B}|+\frac{|\textbf{B}|^2+M_2^2}{2|\textbf{p}|}-B^0,\nonumber\\
   E_3&=|\textbf{p}|-|\textbf{B}|+\frac{|\textbf{B}|^2+M_3^2}{2|\textbf{p}|}+B^0,\nonumber\\
   E_4&=|\textbf{p}|-|\textbf{B}|+\frac{|\textbf{B}|^2+M_4^2}{2|\textbf{p}|}+B^0.
   \label{7.15b}
\end{align}
\begin{figure}[!ht]
    \centering
    \includegraphics[width=8cm]{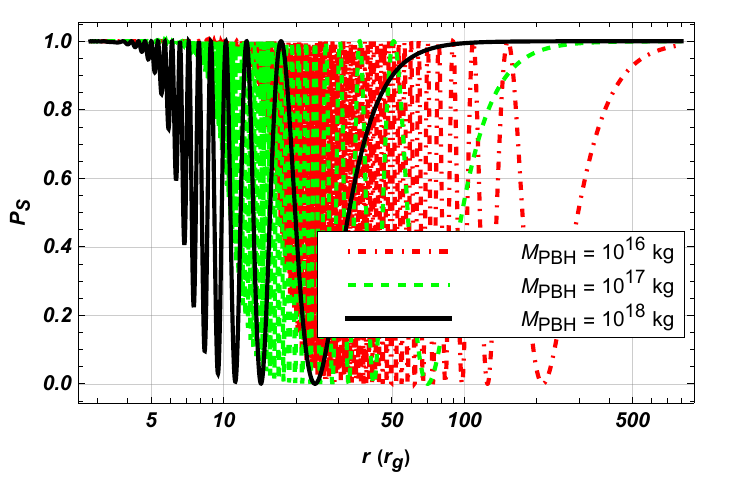}
   \caption{\justifying{In the two-flavor scenario, the survival probability $P_S$ as a function of the radial distance $r$ for different values of the PBH mass, with $a=0.998$ and $\theta=\pi/4$.}}
    \label{Fig14}
\end{figure}Thus, by using Eq.\,(\ref{7.15b}) in Eq.\,(\ref{7.16}), one can determine that at extremely long radial distances ($ r>>100 r_g$) from the PBH, the effect of the four-vector gravitational potential ($B^0$ and $|\textbf{B}|$) diminishes, while the contributions from  $M_1$, $M_2$, $M_3$ and $M_4$ become more dominant. At such long radial distances ($r>>100 r_g$), the different values of $\theta$ and $a$ have the same effect on $P_S$, as shown in the top and bottom panels of Fig.\,{\ref{Fig13}}, respectively. This leads to changes in $P_S$ during the two-flavor oscillation of the neutrino-antineutrino system in vacuum (see the behavior of $P_s$ at $r\geq 10^4 r_g$ in Fig.\,{\ref{Fig13}}). This behavior of $P_s$ at large $r$ resembles the experimentally well-established neutrino oscillations in flat spacetime, which is to be expected, far from the PBH. These results indicate that the two-flavor oscillation of the neutrino-antineutrino system is also influenced by the varying magnitudes of the gravitational scalar ($B^0$) and vector ($|\textbf{B}|$) potentials. However, particularly at long distances ($r>>100 r_g$), no oscillation occurs in the neutrino-antineutrino system (see Fig.\,\ref{Fig7}).

Moreover, using Eq.\,(\ref{7.15a}) in Eq.\,(\ref{7.16}), Fig.\,\ref{Fig14} illustrates $P_S$ of the initial electron flavor neutrino state $\ket{\psi_e}$ as a function of the radial distance $r$ from the PBH for different values of the PBH mass, with $a=0.998$ and $\theta=\pi/4$. We observe that for a low PBH mass, $M_{\text{PBH}}=10^{16}\text{kg}$ (red dot-dashed line), $P_S$ fluctuates more frequently between 1 and 0, exhibiting a highly oscillatory nature. However, for a high PBH mass, $M_{\text{PBH}}=10^{18}\text{kg}$ (black solid line), $P_S$ fluctuates less frequently between 1 and 0, exhibiting a less oscillatory nature. This is because a higher PBH mass corresponds to less curvature, resulting in a weaker spin-curvature coupling. As a result, the oscillatory behavior of $P_S$ during the two-flavor oscillation of the neutrino-antineutrino system is reduced at high PBH mass. However, we observe $P_S$ becoming 1 very close ($r\approx3r_g$) and far ($r\approx800r_g$) from the PBH for all values of the PBH mass.

These results demonstrate that the gravitational effect on the two-flavor oscillations of the neutrino-antineutrino system depends not only on the angle of the position vector of the neutrino spinor with respect to the spin axis of the PBH, but also on the strength of its specific angular momentum as well as the PBH mass, which is significant only over a short range of radial distances ($r\approx 100 r_g$) from the PBH.

\section{QSLT for the two-flavor oscillation of neutrino-antineutrino system in curved spacetime}
\label{sec8}
\begin{figure}[!htbp]
   \centering
    \begin{subfigure}{0.99\columnwidth}
       \centering
        \caption{}
        \label{Fig15a}
       \includegraphics[width=\textwidth]{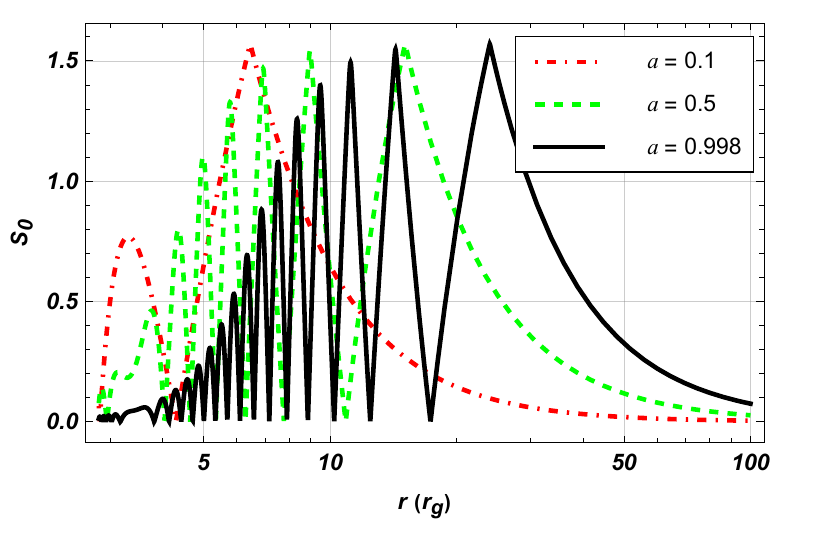}
    \end{subfigure}
   \hfill
   \begin{subfigure}{0.99\columnwidth}
        \centering
        \caption{}
        \label{Fig15b}
        \includegraphics[width=\textwidth]{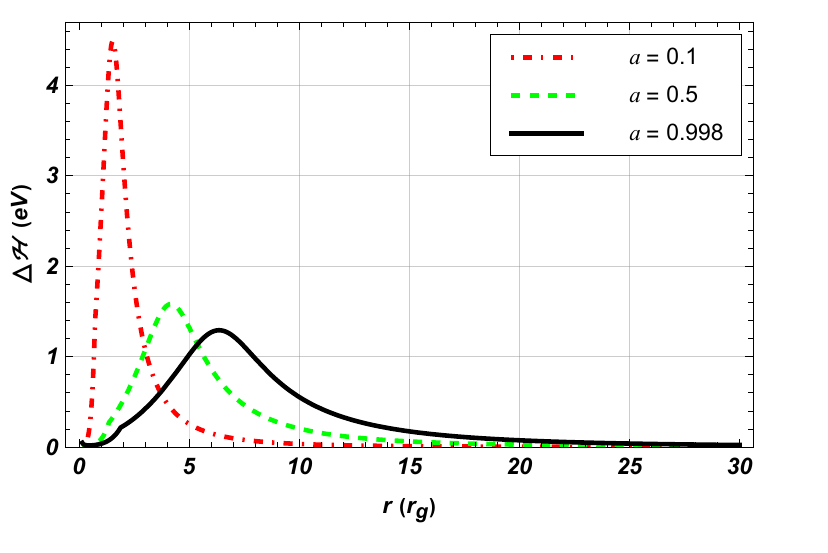} 
    \end{subfigure}
    \hfill
   \begin{subfigure}{0.99\columnwidth}
        \centering
         \caption{}
        \label{Fig15c}
        \includegraphics[width=\textwidth]{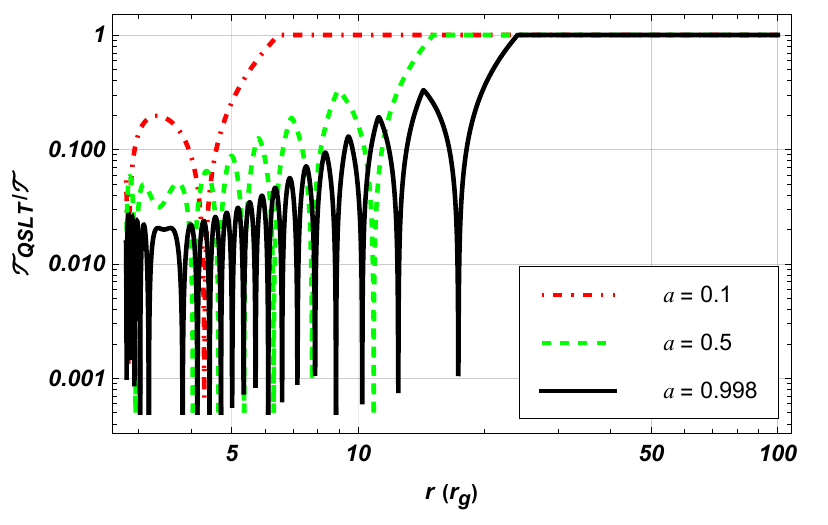}
    \end{subfigure}
   \caption{\justifying{In the two-flavor scenario, (a) the Bures angle $S_0$ (top panel); (b) variance $\Delta{\mathcal{H}}$ (middle panel); and (c) the ratio ${\mathcal{T}_{\text{QSLT}}}/{\mathcal{T}}$ (bottom panel), of the initial state $\ket{\psi_e}$ as a function of the radial distance $r$ for different values of $a$.}}
    \label{Fig15}
\end{figure}

In general, the Dirac Hamiltonian, which can be derived from Eq.\,({\ref{1.8}}), is Hermitian and time independent. Thus, we can apply the concept of quantum speed limit time (QSLT) to the gravity-induced two-flavor oscillations of the neutrino-antineutrino system \cite{Thakuria:2022taf}. If an observer is at infinity, the time-evolved electron flavor neutrino state $\ket{\psi_e(t)}$, given by Eq.\,({\ref{7.9}}), can be reexpressed as a function of radial distance $r$ as
\begin{equation}
   \ket{\psi_e(r)}=T_{e e^c}(r)\ket{\psi^c_e} +T_{e \mu^c}(r)\ket{\psi^c_\mu}
   + T_{ee}(r) \ket{\psi_e}+T_{e \mu}(r) \ket{\psi_\mu}.
   \label{8.1}
\end{equation}
For the state $\ket{\psi_e(r)}$, the Bures angle given by Eq.\,(\ref{a.3}) can be used to measure the distance between its initial and final states, in the Hilbert space, as follows
    \begin{align}
    \mathcal{S}_0(r)&= \cos^{-1}(|\langle \psi_e (0)|\psi_e(r\rangle|)\nonumber\\
    &=\cos^{-1}(\sqrt{P_S(r)})=\cos^{-1}(\sqrt{|T_{ee}(r)|^2}).
    \label{8.2}
    \end{align}
In the top panel of Fig.\,{\ref{Fig15}}, for the initial state $\ket{\psi_e}$, we analyze the Bures angle $S_0$ as a function of the radial distance $r$ from the PBH for different values of $a$, in the ultrarelativistic approximation, with $\theta=\pi/4$.  $S_0$ is suppressed for a small radial distance $r$ from the PBH due to the high strength of the four-vector gravitational potential ($B^0$ and $|\textbf{B}|$) for all values of $a$ (see the bottom panels of Fig.\,{\ref{Fig2}} and  Fig.\,{\ref{Fig3}}). However, due to the low strength of the four-vector gravitational potential at large radial distances $r$, $S_0$ reaches its highest value, $S_0\approx\pi/2$. Since the survival probability $P_s$ changes between 1 and 0 depending on the strength of $a$ (see the bottom panel of Fig.\,{\ref{Fig13}}), $S_0$ given by Eq.\,(\ref{8.2}) exhibits the expected behavior in Fig.\,\ref{Fig15} and oscillates differently for the respective values of $a$. At $r\approx 100 r_g$, since $P_S\rightarrow 1$, $S_0$ tends to minimum. However, for $r>100r_g$, $P_S$ becomes indifferent for all values $a$, as shown in the bottom panel of Fig.\,\ref{Fig13}. Consequently, $S_0$ will also follow similar trends for $r>100r_g$. Thus, when $P_S$ is maximum (or minimum), $S_0$ is minimum (or maximum). It is worth mentioning that, near the PBH, the value of $S_0$ is minimum, which means that the initial and final states are the same. However, as we move further from the PBH the distance between initial and final states starts growing.

Furthermore, using Eq.\,({\ref{8.1}}) in Eq.\,({\ref{a.4}}), we calculate the energy fluctuation $\Delta\mathcal{H}$ for the initial state $\ket{\psi_e}$ as
\begin{equation}
      \Delta{\mathcal{H}}(r)\equiv \sqrt{ [\langle \dot{\psi_e} (r)|\dot{\psi_e}(r)\rangle-(i\langle \psi_e (r)|\dot{\psi_e}(r)\rangle)^2 ]}=\sqrt{\mathcal{Q}(r)},
      \label{8.3}
\end{equation}

\begin{figure}[!htbp]
   \centering
    \begin{subfigure}{0.99\columnwidth}
       \centering
       \caption{}
        \label{Fig16a}
       \includegraphics[width=\textwidth]{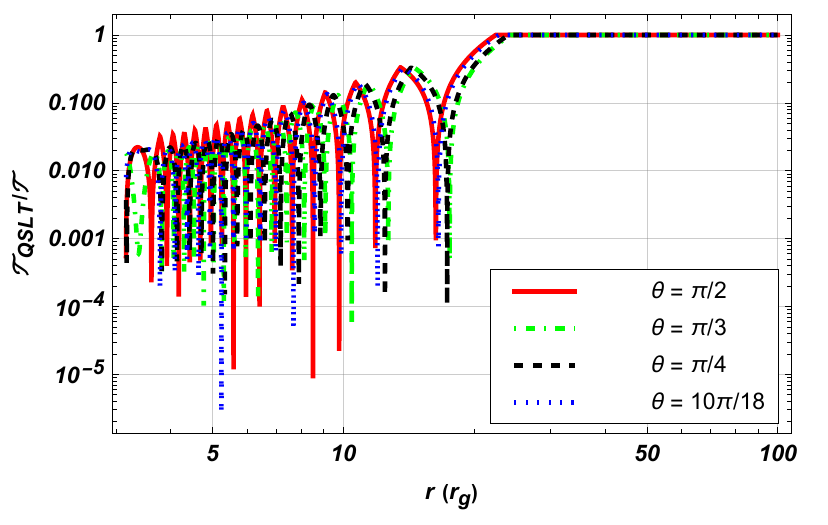}
    \end{subfigure}
   \hfill
   \begin{subfigure}{0.99\columnwidth}
        \centering
        \caption{}
        \label{Fig16b}
        \includegraphics[width=\textwidth]{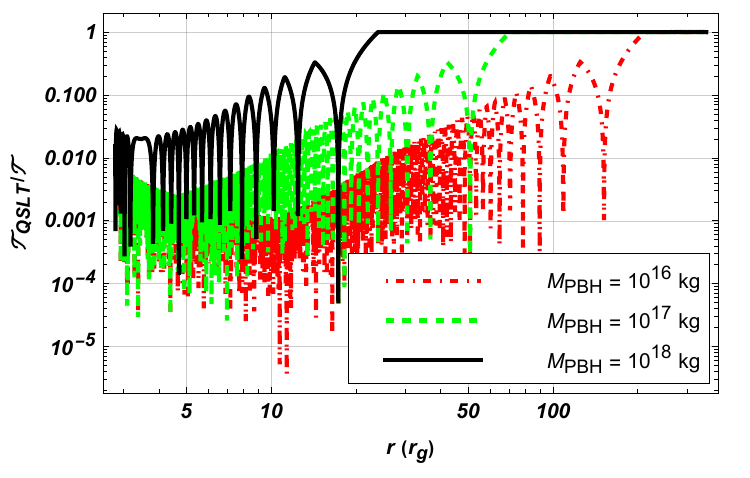}
    \end{subfigure}
   \caption{\justifying{In the two-flavor scenario, the ratio ${\mathcal{T}_{\text{QSLT}}}/{\mathcal{T}}$ of the initial state $\ket{\psi_e}$ as a function of the radial distance $r$: (a) in the top panel, for different values of angle $\theta$, with $M_{\text{PBH}}=10^{18}\text{kg}$; and (b) in the bottom panel, for different values of the PBH mass, with $a=0.998$ and $\theta=\pi/4$.}}
    \label{Fig16}
\end{figure}
where
\begin{align}
    \mathcal{Q}(r)&=[-\frac{1}{16} (E_1 + E_2 + E_3 + E_4 +2 (E_3 - E_4)\cos(2\phi_2) \nonumber\\
    &\cos^2\theta_e + (E_3 + E_4-E_1-E_2)\cos(2\theta_e) + 2 (E_1 - E_2)\nonumber\\
 & \cos(2\phi_1) \sin^2\theta_e)^2 +\frac{1}{4} (E^2_1 + E^2_2 + E^2_3 + 
     E^2_4 +2(E^2_3 - E^2_4)\nonumber\\
   & \cos(2\phi_2)\cos^2\theta_e +(E^2_3+E^2_4-E^2_1-E^2_2) \cos(2\theta_e) \nonumber\\
    & +2 (E^2_1 - E^2_2)\cos(2\phi_1) \sin^2\theta_e)].
    \label{8.4}
\end{align}
In the ultrarelativistic regime, the middle panel of Fig.\,{\ref{Fig15}} illustrates the energy fluctuation $\Delta\mathcal{H}$ as a function of the radial distance $r$ for different values of $a$, with $\theta=\pi/4$. We observe that at small radial distances $r$ from the PBH, $\Delta \mathcal{H}$ is high, and as we increase $r$, after a certain range, $\Delta\mathcal{H}$ tends to zero for all values of $a$. For $r>6 r_g$, the energy fluctuation remains consistently high for $a=0.998$ (black solid line) compared to other values of $a$. 

Finally, by substituting Eq.\,({\ref{8.2}}) and Eq.\,({\ref{8.3}}) in Eq.\,({\ref{a.2}}), and using Eq.\,(\ref{5.6}), we compute the QSLT bound ratio (${\mathcal{T}_{\text{QSLT}}}/{\mathcal{T}}$) of the initial state $\ket{\psi_e}$ as 
\begin{equation}
    \frac{\mathcal{T}_{\text{QSLT}}(r)}{\mathcal{T}(r)}=\frac{\left[\frac{\cos^{-1}(\sqrt{|T_{ee}(r)|^2})}{\sqrt{\mathcal{Q}(r)}}\right]}{\int^r_{r_1> r_{+}}f(r,\theta,a)dr},
    \label{8.5} 
\end{equation}
where $\hbar=1$, and $t=\mathcal{T}.$
 In the ultrarelativistic limit, the bottom panel of Fig.\,{\ref{Fig15}} illustrates the ratio $\mathcal{T}_{\text{QSLT}}/\mathcal{T}$ of the initial state $\ket{\psi_e}$ as a function of the radial distance $r$ for different values of $a$, with $\theta=\pi/4$. We find the following empirical relation between them,
\begin{equation}
    \left(\frac{\mathcal{T}_{\text{QSLT}}(r)}{\mathcal{T}(r)}\right)_{a=0.998}< \left(\frac{\mathcal{T}_{\text{QSLT}}(r)}{\mathcal{T}(r)}\right)_{a=0.5}< \left(\frac{\mathcal{T}_{\text{QSLT}}(r)}{\mathcal{T}(r)}\right)_{a=0.1}.
    \label{8.6}
\end{equation}
 We observe that at low $a$ (red dot-dashed line), the ratio ${\mathcal{T}_{\text{QSLT}}}/{\mathcal{T}}$ behaves in a tight manner and quickly approaches 1, compared to other values of $a$. It suggests that the minimum time required for oscillation is maximized at $a=0.1$ (red dot-dashed line). Conversely, as $a$ increases to higher values, such as $a=0.998$ (black solid line), the ratio ${\mathcal{T}_{\text{QSLT}}}/{\mathcal{T}}$ decreases, indicating that the initial state $\ket{\psi_e}$ oscillates more rapidly for a longer range of $r$. This behavior is expected because a large specific angular momentum ($a=0.998$) corresponds to high strength of $B^0$ and $|\textbf{B}|$ (see the magenta dotted line in the bottom panels of Fig.\,{\ref{Fig2}} and Fig.\,{\ref{Fig3}}), leading to minimum value of ${\mathcal{T}_{\text{QSLT}}}/{\mathcal{T}}$ throughout the range of $r$ (see the black solid line in the lower panel of Fig.\,{\ref{Fig15}}), compared to lower values of $a$.  Hence, it is evident from Eq.\,({\ref{8.6}}) that the ratio ${\mathcal{T}_{\text{QSLT}}}/{\mathcal{T}}$ for the two-flavor oscillation of neutrino-antineutrino system in curved spacetime is significantly suppressed by increasing $a$. However, it is important to note that these results are significant only within a short range of radial distances from the spinning PBH. At long radial distances ($r>30 r_g$), the gravitational effects diminish, leading to ${\mathcal{T}_{\text{QSLT}}}/{\mathcal{T}}\rightarrow1$ across all values of $a$. This implies a slower dynamical evolution of the initial state $\ket{\psi_e}$, as it achieves the maximum QSLT bound ratio.

 Moreover, similar to the previous calculation, we use Eq.\,({\ref{8.5}}) to compute the ratio ${\mathcal{T}_{\text{QSLT}}}/{\mathcal{T}}$ for different values of $\theta$ with $a=0.998$. In the top panel of Fig.\,{\ref{Fig16}}, we examine the behavior of the ratio ${\mathcal{T}_{\text{QSLT}}}/{\mathcal{T}}$ as a function of the radial distance  $r$ for different values of $\theta$. We observe that for all values of $\theta$, the ratio ${\mathcal{T}_{\text{QSLT}}}/{\mathcal{T}}<1$ for a short range of $r$, which signifies a faster dynamical evolution of the initial quantum state $\ket{\psi_e}$ in this regime.

 Finally, using Eq.\,({\ref{8.5}}), we compute the ratio ${\mathcal{T}_{\text{QSLT}}}/{\mathcal{T}}$ as a function of the radial distance $r$ for different values of the PBH mass, with $a=0.998$ and $\theta=\pi/4$. The empirical relation we obtain is as follows
\begin{align}
    \left(\frac{\mathcal{T}_{\text{QSLT}}(r)}{\mathcal{T}(r)}\right)_{M_{\text{PBH}}=10^{16} \text{kg}}&< \left(\frac{\mathcal{T}_{\text{QSLT}}(r)}{\mathcal{T}(r)}\right)_{M_{\text{PBH}}=10^{17} \text{kg}}\nonumber\\
    &< \left(\frac{\mathcal{T}_{\text{QSLT}}(r)}{\mathcal{T}(r)}\right)_{M_{\text{PBH}}=10^{18} \text{kg}}.
    \label{8.7}
\end{align}
 This result indicates that nearer to PBH, a low PBH mass, $M_{\text{PBH}}=10^{16}\text{kg}$ (red dot-dashed line), corresponds to a faster dynamical evolution of the initial quantum state $\ket{\psi_e}$, due to a higher spin-curvature coupling.
 \section{Entanglement in two-flavor oscillations of the neutrino-antineutrino system in curved spacetime}
 \label{sec9}
 
Now, we treat the two-flavor oscillations of the neutrino-antineutrino system as a four-qubit system under the impact of neutrino-antineutrino mixing caused by the gravitational field. 
The representation of the occupation number can be expressed as \cite{Dixit:2019lsg}
\begin{eqnarray}
   { \ket{\psi_e^c}\equiv \ket{1}_{e^c}\otimes\ket{0}_{\mu^c}\otimes \ket{0}_{e}\otimes\ket{0}_\mu\equiv\ket{1000}},&\nonumber\\
   { \ket{\psi_\mu^c}\equiv \ket{0}_{e^c}\otimes\ket{1}_{\mu^c}\otimes \ket{0}_{e}\otimes\ket{0}_\mu\equiv\ket{0100}},&\nonumber\\
   { \ket{\psi_e}\equiv \ket{0}_{e^c}\otimes\ket{0}_{\mu^c}\otimes \ket{1}_{e}\otimes\ket{0}_\mu\equiv\ket{0010}},&\nonumber\\
   { \ket{\psi_\mu}\equiv \ket{0}_{e^c}\otimes\ket{0}_{\mu^c}\otimes \ket{0}_{e}\otimes\ket{1}_\mu\equiv\ket{0001}}.&\nonumber\\
   \label{9.1}
\end{eqnarray}
\begin{figure}[!ht]
    \centering
    \includegraphics[scale=0.7]{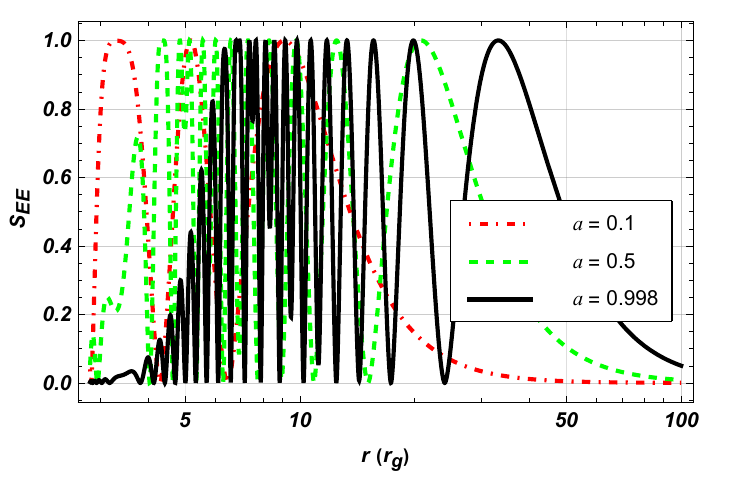}
    \caption{\justifying{In the two-flavor scenario, the entanglement entropy $S_{EE}$ of the initial state $\ket{\psi_e}$ as functions of radial distance $r$ for different values of $a$, when $\theta=\pi/4$.}}
    \label{Fig17}
\end{figure}
Thus, using eqn.\,(\ref{9.1}) in eqn.\,(\ref{8.1}), the time evolved electron flavor neutrino state as a function of $r$ in the four-qubit system can be written as
  \begin{eqnarray}
   {\ket{\psi_e(r)}=T_{e e^c}(r)\ket{1000}+T_{e \mu^c}(r)\ket{0100}}&\nonumber\\
   {+ T_{ee}(r) \ket{0010}+T_{e \mu}(r) \ket{0001}}&.
   \label{9.2}
\end{eqnarray}  
For unitary dynamics, the simplest case of a multipartite quantum system is a bipartite pure quantum system \cite{Guhne:2008qic}. An example of a bipartite pure quantum system is Bell’s state. If a $16\times16$ density matrix  $\rho(r)=\ket{\psi_e(r)}\bra{\psi_e(r)}$ of the state $\ket{\psi_e(r)}$ consists of two subsystems $\rho_e(r)$ and $\rho_{({e^c}\mu{\mu^c})}(r)$, entanglement in the bipartite pure quantum system implies
\begin{equation}
    \rho(r)\neq\rho_e(r)\otimes\rho_{({e^c}\mu{\mu^c})}(r),
    \label{9.3}
\end{equation}
where $\rho_e(r)=\rm Tr_{({e^c}\mu{\mu^c})}[\rho(r)]$ and $\rho_{({e^c}\mu{\mu^c})}(r)=\rm Tr_e[\rho(r)]$ are two reduced density matrices. These reduced-density matrices share the same set of eigenvalues. It is worth mentioning that due to the unitary dynamics, $\rho(r)$ always exists in a pure quantum state with properties such as $\rho(r)=\rho^2(r)$ (idempotent matrix) and $\mathrm{Tr}[\rho^2(r)]=1$. The reduced density matrices $\rho_e(r)$ and $\rho_{({e^c}\mu{\mu^c})}(r)$ follow $ \mathrm{Tr}[\rho^2_e(r)]<1$ and $\mathrm{Tr}[\rho^2_{({e^c}\mu{\mu^c})}(r)]<1$, indicating mixed states. 

Moreover, the entanglement entropy ($S_{EE}(r)$) serves as a crucial quantum correlation to assess the strength of quantum entanglement between the subsystems of a bipartite quantum system. In the context of the two-flavor oscillations of the neutrino-antineutrino system in curved spacetime, $S_{EE}(r)$ provides valuable insights into how entanglement evolves with the radial distance from the spinning PBH. The entanglement entropy $S_{EE}(r)$, determined by the Von Neumann entropy on the reduced density matrix of one of the subsystems, is defined by \cite{Shrimali:2022bvt,Bennett:1995tk,PhysRevA.67.012313,2004IJTP...43.1241P}
\begin{equation}
   S_{EE}(r)=S(\rho_e(r))=-\mathrm{Tr}[\rho_e(r) \rm log_2\rho_e(r)].
   \label{9.4}
\end{equation}
The required $2\times2$ reduced density matrix of the state $\ket{\psi_e(r)}$ can be expressed as
\begin{equation}
 \rho_{e}(r)=\text{diag}(\lambda_1(r),\lambda_2(r)),
 \label{9.5}
\end{equation}
where $\lambda_1(r)$ and $\lambda_2(r)$ are the two nonzero eigenvalues of $\rho_e(r)$. It is important to note that other combinations of reduced density matrices are also possible for $\rho(r)$ such as $\rho_\mu(r)$, $\rho_{e^c}(r)$, $\rho_{\mu^c}(r)$, but they will also give two nonzero eigenvalues. By using $\rho_e(r)$ given by Eq.\,(\ref{9.5}) in Eq.\,(\ref{9.4}), we find the expression of the entanglement entropy $ S_{EE}(r)$ for the initial state $\ket{\psi_e}$ as
\begin{equation}
    S_{EE}(r)=-\lambda_1(r) \rm log_2\lambda_1(r)-\lambda_2(r) \rm log_2\lambda_2(r).
    \label{9.6}
\end{equation}

\begin{figure}[!htbp]
   \centering
    \begin{subfigure}{0.99\columnwidth}
       \centering
        \caption{}
        \label{Fig18a}
       \includegraphics[width=\textwidth]{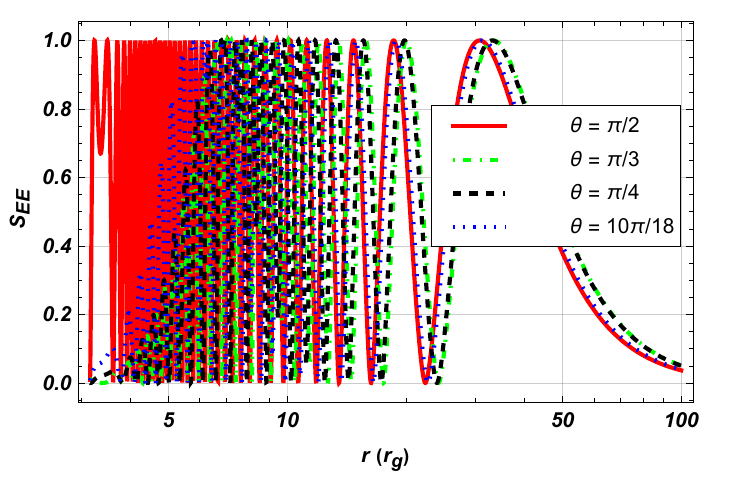}
    \end{subfigure}
   \hfill
   \begin{subfigure}{0.99\columnwidth}
        \centering
        \caption{}
        \label{Fig18b}
        \includegraphics[width=\textwidth]{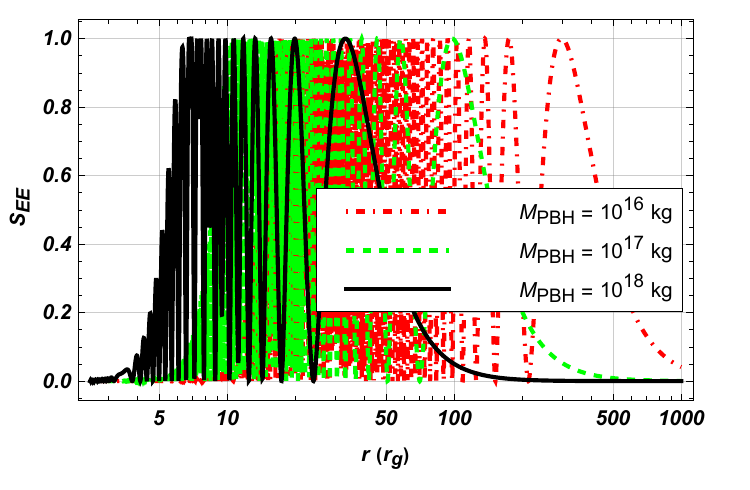}
    \end{subfigure}
   \caption{\justifying{In the two-flavor scenario, the entanglement entropy $S_{EE}$ of the initial state $\ket{\psi_e}$ as a function of the radial distance $r$: (a) in the top panel, for different values of angle $\theta$, with
   $M_{\text{PBH}}=10^{18}\text{kg}$; and (b) in the bottom panel, for different values of the PBH mass, with $a=0.998$ and $\theta=\pi/4$.}}
    \label{Fig18}
\end{figure}

Using Eq.\,(\ref{9.6}), Fig.\,\ref{Fig17} illustrates the entanglement entropy ($S_{EE}$) of the initial state $\ket{\psi_e}$ as a function of the radial distance $r$ from the PBH, for different values of $a$. At radial distance $r\approx r_+$, $\lambda_1(r)$ and $\lambda_2(r)$ approach 0 and 1, respectively, leading to vanishing $S_{EE}$. For $r>r_+$, $S_{EE}>0$, indicating the presence of bipartite entanglement in the two-flavor oscillations of the neutrino-antineutrino system in curved spacetime. We observe that at a small radial distance $r$, the $S_{EE}$ decreases as $a$ increases from a low (red dot-dashed line) to a high value (black solid line). It occurs because a high value of $a$ (black solid line) corresponds to the high strength of the four-vector gravitational vector potential ($B^0$ and $|\textbf{B}|$), which causes the initial
flavor state $\ket{\psi_e}$ to stabilize. However, since the strength of $B^0$ and $|\textbf{B}|$ decreases with increasing radial distance $r$ for all $a$, $S_{EE}$ increases according to the value of $a$. Another reason for the high strength of $S_{EE}$ for a low value of $a$ (red dot-dashed line) near the PBH is energy fluctuation $\Delta H$, which is very high in this regime (see the red dot-dashed line in the middle panel of Fig.\,{\ref{Fig15}}). Moreover, a classic example of a maximally pure bipartite entangled state is Bell’s state. In the bottom panel of Fig.\,{\ref{Fig13}}, whenever the survival probability ($P_s$) of the initial neutrino flavor state $\ket{\psi_e}$ is around $0.5$, the system is in a bipartite Bell-state-like configuration, which is maximally entangled for all values of $a$. This is why $S_{EE}\rightarrow1$ at these points in Fig.\,\ref{Fig17} for all values of $a$. Moreover, at $r\approx 100 r_g$, $P_S\rightarrow 1$ for all $a$. Consequently,  $S_{EE}$ becomes minimum and indifferent across all values of $a$. 

Furthermore, using Eq.\,(\ref{9.6}), the top panel of Fig.\,\ref{Fig18} depicts $S_{EE}$ of the initial state $\ket{\psi_e}$ as a function of the radial distance $r$ from the PBH, for different values of $\theta$, with $a=0.998$. We observe that at a small radial distance $r$, $S_{EE}$ is maximum at $\theta=\pi/2$ (red solid line) and minimum at other values of $\theta$. This is because the strength of $B^0$ is low at $\theta=\pi/2$, compared to other values of $\theta$. However, at large radial distances $r$, the strength of $B^0$ and $|\textbf{B}|$ decrease for all $\theta$. Consequently, $S_{EE}$ increases for all values of $\theta$. Whenever, $P_S$ crosses the value $0.5$ (see the top panel of Fig.\,{\ref{Fig13}}), the $S_{EE}$ becomes maximum, with the initial state $\ket{\psi_e}$ exhibiting a maximally pure bipartite entangled state for all values of $\theta$. At $r\approx100r_g$, since $P_S$ becomes indifferent to all $\theta$, therefore $S_{EE}$ also becomes indifferent to all $\theta$. Hence, within a short range of radial distances $r$, the entanglement entropy in the two-flavor oscillations of the neutrino-antineutrino system in curved spacetime is significantly suppressed due to the high strength of the four-vector gravitational potential.

Moreover, using Eq.\,(\ref{9.6}), the bottom panel of Fig.\,\ref{Fig18} depicts $S_{EE}$ of the initial state $\ket{\psi_e}$ as a function of the radial distance $r$ from the PBH, for different values of the PBH mass, with $a=0.998$, and $\theta=\pi/4$. We observe that similar to the $P_S$ behavior in the Fig.\,\ref{Fig14}, $S_{EE}$ for a low PBH mass, $M_{\text{PBH}}=10^{16}\text{kg}$ (red dot-dashed line), exhibits a maximally bipartite pure entangled state that is highly oscillatory in nature, compared to a high PBH mass. However, the fact that $S_{EE}$ is suppressed near the PBH and increases farther away from the PBH holds true for all different values of PBH mass. At $r\approx1000r_g$, $S_{EE}$ becomes indifferent to all $\theta$.

\section{Conclusions}
\label{sec10}
In this work, we have considered a rotating black hole described by the Kerr metric in Kerr-Schild polar (KSP) coordinates. We found an analytical expression for the four-vector gravitational potential with the Hermitian Dirac Hamiltonian using these coordinates. This gravitational potential results in an axial vector term in the Dirac equation in curved spacetime. Our findings indicate that the magnitude of the gravitational vector potential is highly dependent on both the angle of the position vector of the neutrino spinor with respect to the spin axis of the PBH and the strength of the specific angular momentum ($a$). We observe a significant increase in the magnitude of the gravitational vector potential at high values of $a$, particularly at $\theta=\pi/3$. This four-vector gravitational potential modifies the mass matrix of the neutrino-antineutrino system and gives the gravitational Zeeman effect (GZE). For large $a$, the gravitational potential is large, leading the mixing angle for the neutrino-antineutrino system to decrease to a low value at small radial distances from the PBH. Conversely, for a large radial distance from the PBH, the gravitational potential weakens, and the mixing angle saturates at its maximum value of $\pi$/4 for all values of $a$ in the neutrino-antineutrino system. Similar results were also obtained for the mixing angles with a fixed value of $a$ and for different values of $\theta$. We also investigated the transition probability during the evolution of the neutrino-antineutrino system and for the two-flavor oscillation of the neutrino-antineutrino system in curved spacetime. In the ultrarelativistic limit, from the perspective of an observer at infinity, the transition probability of the initial flavor neutrino state varies significantly depending on the mass of black hole $M_{\text{PBH}}$, coordinates such as $\theta$, and $a$. These effects are noticeable within a short radial distance from the spinning PBH. For extremely larger radial distances, the strength of the four-vector gravitational potential diminishes, leading to two-flavor oscillations of the neutrino-antineutrino system in vacuum for all values of $\theta$ and $a$. 

In the presence of strong gravitational fields near the PBH, the survival probability is high. Due to the strong spin-gravity coupling, neutrinos are less susceptible to oscillate into their corresponding flavor state. This effect is evident for all neutrino trajectories and the different spin parameter values considered in this study, which further showcases the importance of gravitational coupling to neutrinos' spin in its flavor transitions.

Additionally, we have examined the Bures angle for the initial neutrino flavor state in the two-flavor oscillation of the neutrino-antineutrino system as it evolves with radial distance from the PBH. Our results show that the distance between the initial and final flavor states changes with $a$ of the PBH. We have also studied how the energy fluctuation of the initial neutrino flavor state evolves with the radial distance from the PBH and observed that the energy fluctuation is higher for a larger $a$. Using these concepts, we estimated the quantum speed limit time bound ratio (${\mathcal{T}_{\text{QSLT}}}/{\mathcal{T}}$) of the two-flavor oscillation of the neutrino-antineutrino system in curved spacetime. Our observations reveal that for short radial distance from the PBH, faster flavor transition of the initial neutrino flavor state is possible for large values of $a$ of the PBH when the gravitational potential is high. We have also observed that a low PBH mass ($M_{\text{PBH}}$) corresponds to a faster dynamical evolution of the initial flavor state and vice versa.

Furthermore, the entanglement entropy of the initial neutrino flavor state as it varies with radial distance from the spinning PBH was studied. It was observed that the two-flavor oscillation of the neutrino-antineutrino system in curved spacetime exhibits bipartite pure entanglement and resembles a Bell-state configuration. In strong gravity, a large $a$ leads to significant suppression of entanglement during the two-flavor oscillation of the neutrino-antineutrino system within a short radial distance ($r\leq 6\,r_g$) from the PBH, owing to the magnitude of survival probability being less than 0.5. This result also holds true when the entanglement entropy varies with radial distance for different values of $\theta$ and the PBH mass. However, we have reported that a low PBH mass corresponds to entanglement entropy that is highly oscillatory in nature compared to a high PBH mass.

Moreover, depending on the spin-curvature coupling (i.e. GZE), the neutrino and antineutrino will have different energy. Considering these energies, one can further take account of statistical distributions and deduce the asymmetry of number density for neutrinos. From statistical calculations, the number density of neutrinos can be quantified as, $n_\nu\sim gT^3(3\zeta(3)/2-\pi^2B^0/6T)$, where $g=2$ is the gyromagnetic ratio, $T$ is the temperature of the given system, and $\zeta(.)$ is the Riemann zeta function. The corresponding neutrino number density asymmetry is, $\Delta n=(n_\nu-n_{\nu^c})\sim g\pi^2T^3(B^0/3T)$ \cite{Mukhopadhyay:2007vca,Singh:2003sp,Mukhopadhyay:2005gb,Debnath:2005wk}. The neutrino number asymmetry per unit neutrino is then given by $\Delta n/n$. For the Sun, this number comes out to be $\sim10^{-39}$ at its surface ($T=5772K$). Comparing this with the observed value $\Delta n/n=10^{-10}$ \cite{Fields:2006ga,Steigman:2010zz,Komatsu_2011,Canetti:2012zc}, the asymmetry around Sun is quite small. A future solar neutrino probe with the required sensitivity will be able to detect the aforementioned neutrino number density asymmetry on the surface. The neutrino asymmetry is a direct consequence of neutrino-antineutrino oscillations due to GZE. A direct/indirect detection of $\Delta n/n$ will further validate curvature coupling of neutrinos.

In conclusion, we have shown that the spin-curvature coupling of the neutrino with the background spacetime leads to significant changes in its dynamics. Oscillatory properties of its various quantum parameters are suppressed in the high coupling regime. This effect reduces as the particle propagates away from the PBH, due to reduced coupling.

\section*{Acknowledgements}
 A.~K.~J., B.~M., and S.~B. would like to acknowledge the project funded by SERB, India, with Ref. No. CRG/2022/003460, for supporting this research. M.~D. acknowledges the financial support of the Ministry of Education (MoE) fellowship scheme. M.~P. gratefully thanks the Prime Minister's Research Fellows (PMRF) scheme for providing fellowship. 
A.~K.~J. and M.~D. thank Ranjan Laha (IISc) for helpful discussions. 

\section*{Data Availability}
The data that support the findings of this article are openly available~\cite{ParticleDataGroup:2024cfk}.
\bibliography{reference.bib}

\end{document}